\documentclass[sigconf,9pt]{acmart}

\usepackage[english]{babel}
\usepackage{blindtext}
\usepackage{tikz,pgfplots}
\usetikzlibrary{calc}
\usepackage{tkz-euclide}
\pgfplotsset{compat=1.5}
\usepgfplotslibrary{polar}
\usepackage{epstopdf} 
\usepackage{amsmath}
\usepackage{wrapfig}
\usepackage{verbatim}
\usepackage{mathrsfs}
\usepackage{accents}
\usepackage{acronym}
\usepackage{graphicx}
\usepackage{textcomp}
\usepackage{xcolor}
\usepackage{mathtools}
\usepackage{color,soul}
\usepackage{graphicx}
\usepackage{booktabs}
\usepackage[font={small}]{caption}
\usepackage[font={small}]{subcaption}
\usepackage{algorithm}
\usepackage{algorithmic}
\usepackage{fancyhdr}

\DeclareMathOperator*{\argmin}{arg\,min}

\usepackage{caption}
\usepackage{subcaption}
\usepackage{cleveref}
\usepackage{url}

\newcommand{\eq}[1]{Eq.~\eqref{#1}}
\newcommand{\fig}[1]{Fig.~\ref{#1}}
\newcommand{\tab}[1]{Tab.~\ref{#1}}
\newcommand{\secref}[1]{Section~\ref{#1}}
\newcommand{\alg}[1]{Alg.~\ref{#1}}

\acrodef{prop}[\textit{MIMORPH}]{MIMO Radio Platform for Heterogeneous wireless systems}

\acrodef{abft}[A-BFT]{Association Beamforming Training}
\acrodef{ack}[ACK]{Acknowledge}
\acrodef{adc}[ADC]{Analog-to-Digital Converter}
\acrodef{aoa}[AoA]{Angle of Arrival}
\acrodef{aod}[AoD]{Angle of Departure}
\acrodef{ap}[AP]{Access Point}
\acrodef{amc}[AMC]{Advanced Mezzanine Card}
\acrodef{awv}[AWV]{Antenna Wave Vector}
\acrodef{axi}[AXI]{Advanced eXtensible Interface}
\acrodef{ber}[BER]{Bit Error Rate}
\acrodef{bft}[BFT]{Beamforming Training}
\acrodef{bp}[BP]{Beam Pattern}
\acrodef{brp}[BRP]{Beam Refinement Phase}
\acrodef{cs}[CS]{Compressed Sensing}
\acrodef{cdf}[CDF]{Cumulative Distribution Function}
\acrodef{cef}[CEF]{Channel Estimation Field}
\acrodef{cfo}[CFO]{Carrier Frequency Offset}
\acrodef{cir}[CIR]{Channel Impulse Response}
\acrodef{csi}[CSI]{Channel State Information}
\acrodef{csirs}[CSI-RS]{CSI-Reference Signal}
\acrodef{cv}[CV]{Constant Velocity}
\acrodef{cnn}[CNN]{Convolutional Neural Network}
\acrodef{cots}[COTS]{Commercial-Off-The-Shelf}
\acrodef{dl}[DL]{Deep Learning}
\acrodef{dma}[DMA]{Direct Memory Access}
\acrodef{dmg}[DMG]{Directional Multi Gigabit}
\acrodef{dti}[DTI]{Data Transfer Interval}
\acrodef{edmg}[EDMG]{Enhanced Directional Multi Gigabit}
\acrodef{ekf}[EKF]{Extended Kalman Filter}
\acrodef{elu}[ELU]{Exponential-Linear Unit}
\acrodef{fmcw}[FMCW]{Frequency-Modulated Continuous-Wave}
\acrodef{fov}[FOV]{Field-of-View}
\acrodef{ft}[FT]{Fourier Transform}
\acrodef{gpio}[GPIO]{General Purpose Input/Output}
\acrodef{gsps}[GSPS]{Giga-Samples per Second}
\acrodef{har}[HAR]{Human Activity Recognition}
\acrodef{ht}[HT]{High Throughput}
\acrodef{if}[IF]{Intermediate Frequency}
\acrodef{ifs}[IFS]{Inter-Frame Spacing}
\acrodef{iht}[IHT]{Iterative Hard Thresholding}
\acrodef{isac}[ISAC]{Integrated Sensing And Communication}
\acrodef{jcr}[JCR]{Joint Communication \& Radar}
\acrodef{jpdaf}[JPDAF]{Joint Probabilistic Data Association Filter}
\acrodef{los}[LOS]{Line-of-Sight}
\acrodef{lbm}[LBM]{Loop-Back Memory}
\acrodef{mae}[MAE]{Mean Absolute Error}
\acrodef{mcs}[MCS]{Modulation and Coding Scheme}
\acrodef{md}[$\mu$D]{micro-Doppler}
\acrodef{mimo}[MIMO]{Multiple Input Multiple Output}
\acrodef{mmwave}[mmWave]{Millimeter-Wave}
\acrodef{msps}[MSPS]{Mega-Samples per Second}
\acrodef{mu}[MU]{Multiple User}
\acrodef{MUSIC}[MUSIC]{MUlti SIgnal Classification}
\acrodef{nac}[NAC]{Normalized Auto Correlation}
\acrodef{nco}[NCO]{Numerical Controlled Oscillator}
\acrodef{nlos}[NLOS]{Non-Line-of-Sight}
\acrodef{ofdm}[OFDM]{Orthogonal Frequency Division Multiplexing}
\acrodef{per}[PER]{Packet Error Rate}
\acrodef{phy}[PHY]{Physical Layer}
\acrodef{pl}[PL]{Programmable Logic}
\acrodef{pov}[POV]{Point-of-View}
\acrodef{ps}[PS]{Processing System}
\acrodef{rf}[RF]{Radio Frequency}
\acrodef{rfsoc}[RFSoC]{Radio Frequency System on a Chip}
\acrodef{rss}[RSS]{Received Signal Strength}
\acrodef{rom}[ROM]{Read Only Memories}
\acrodef{sc}[SC]{Single Carrier}
\acrodef{sdr}[SDR]{Software Defined Radio}
\acrodef{siso}[SISO]{Single Input Single Output}
\acrodef{sls}[SLS]{Sector Level Sweep}
\acrodef{snr}[SNR]{Signal-to-Noise Ratio}
\acrodef{soc}[SoC]{System on a Chip}
\acrodef{spb}[SPB]{Signal Processing Blocks}
\acrodef{srrc}[SRRC]{Square-Root-Raised-Cosine}
\acrodef{ssb}[SSB]{Synchronization Signal Block}
\acrodef{ssr}[SSR]{Super Sample Rate}
\acrodef{sta}[STA]{Station}
\acrodef{stf}[STF]{Short Training Field}
\acrodef{stft}[STFT]{Short Time Fourier Transform}
\acrodef{su}[SU]{Single User}
\acrodef{tf}[TF]{Time-Frequency}
\acrodef{toa}[ToA]{Time of Arrival}
\acrodef{usrp}[USRP]{Universal Software Radio Peripheral}
\acrodef{vht}[VHT]{Very High Throughput}
\acrodef{wlan}[WLAN]{Wireless Local Area Network}

\newcommand{\rev}{\textcolor{black}} 

{
\theoremstyle{plain}

 }

\copyrightyear{2022}
\acmYear{2022}
\acmConference[IPSN]{IEEE/ACM Conference on Information Processing in Sensor Networks}{May 04--06, 2022}{Milan, Italy}
\acmBooktitle{IEEE/ACM Conference on Information Processing in Sensor Networks (IPSN)}
\acmPrice{31.00}
\acmDOI{10.1109/IPSN54338.2022.00014}

\begin{document}
\fancypagestyle{firststyle}
{
    \fancyhf{}
    \chead{2022 21st ACM/IEEE International Conference on Information Processing in Sensor Networks (IPSN)}
    \lfoot{\vspace{5pt} 978-1-6654-9624-7/22/\$31.00 ©2022 \\
    IEEE DOI 10.1109/IPSN54338.2022.0001}
   \renewcommand{\headrulewidth}{0pt} 
}

\title[SPARCS: A Sparse Recovery Approach for Integrated Communication and Human Sensing in mmWave Systems]{SPARCS: A Sparse Recovery Approach for Integrated Communication and Human Sensing in \acs{mmwave} Systems}

\author{Jacopo Pegoraro}
\email{pegoraroja@dei.unipd.it} 
\affiliation{
\institution{University of Padova}
\city{Padova}
\country{Italy}}

\author{Jesus O. Lacruz}
\email{jesusomar.lacruz@imdea.org} 
\affiliation{
\institution{IMDEA Networks Institute}
\city{Madrid}
\country{Spain}}

\author{Michele Rossi}
\email{michele.rossi@unipd.it} 
\affiliation{
\institution{University of Padova}
\city{Padova}
\country{Italy}}

\author{Joerg Widmer}
\email{joerg.widmer@imdea.org} 
\affiliation{
\institution{IMDEA Networks Institute}
\city{Madrid}
\country{Spain}}



\settopmatter{printacmref=false, printccs=false, printfolios=false}

\renewcommand\footnotetextcopyrightpermission[1]{} 

\begin{abstract}
A well established method to detect and classify human movements using \ac{mmwave} devices is the time-frequency analysis of the \mbox{small-scale} Doppler effect (termed micro-Doppler) of the different body parts, which requires a regularly spaced and dense sampling of the \ac{cir}. This is currently done in the literature either using special-purpose radar sensors, or interrupting communications to transmit dedicated sensing waveforms, entailing high overhead and channel utilization.
In this work we present SPARCS, an integrated human sensing and communication solution for \ac{mmwave} systems. SPARCS is the first method that reconstructs high quality signatures of human movement from irregular and sparse \ac{cir} samples, such as the ones obtained during communication traffic patterns. To accomplish this, we formulate the micro-Doppler extraction as a sparse recovery problem, which is critical to enable a smooth integration between communication and sensing. Moreover, if needed, our system can seamlessly inject short \ac{cir} estimation fields into the channel whenever communication traffic is absent or insufficient for the micro-Doppler extraction. SPARCS effectively leverages the intrinsic sparsity of the \ac{mmwave} channel, thus drastically reducing the sensing overhead with respect to available approaches.
We implemented SPARCS on an IEEE~802.11ay \ac{sdr} platform working in the $60$~GHz band, collecting standard-compliant \ac{cir} traces matching the traffic patterns of real WiFi access points. Our results show that the micro-Doppler signatures obtained by SPARCS enable a typical downstream application such as human activity recognition with more than $7$ times lower overhead with respect to existing methods, while achieving better recognition performance.
\end{abstract}

\maketitle
\thispagestyle{firststyle}
\acresetall

\section{Introduction} \label{sec:intro}

There is a growing interest in human tracking \cite{wu2020mmtrack}, \ac{har} \cite{singh2019radhar} and person identification \cite{zhao2019mid} using \ac{mmwave} devices, where the high carrier frequency and large available bandwidth allow for accurate Doppler estimation and precise localization and tracking. To fully exploit these properties, a large body of work has focused on dedicated \ac{mmwave} radars, that adopt specifically designed frequency modulated transmissions to extract the \ac{md} effect caused by human motion (a so-called \ac{md} signature) \cite{vandersmissen2018indoor, seifert2019toward}.

At the same time, given the increasing number of \ac{mmwave} network technologies such as $3$GPP $5$G-NR \cite{5G-NR_R15} and IEEE~802.11ad/ay \cite{802.11ad, 802.11ay}, \ac{isac} solutions are highly appealing. They effectively repurpose communication devices by endowing them with additional environment sensing capabilities, thus avoiding the cost of installing dedicated radar hardware. 
This recent trend has led to the identification of sensing as a key feature of next generation $6$G mobile networks \cite{liu2021integrated} and the creation of the IEEE~802.11bf standardization group \cite{restuccia2021ieee}, aimed at enabling sensing features in \acp{wlan}. However, \ac{isac} system designs are still very limited, focusing on joint communication and sensing waveform design \cite{liu2018toward}, which would require significant modifications to existing communication protocols and a reduction in the achievable communication data rates. Other approaches \cite{liu2021integrated, li2021rethinking, pegoraro2021rapid, zhang2020mmeye} need to alternate communication and sensing phases according to a time-division scheme, where regularly spaced, radar-like transmissions are performed during dedicated sensing periods. This is needed to perceive the fine-grained \ac{md} effect of human motion, for which dense and regular sampling of the \ac{cir} is required, causing significant overhead and channel occupation.

In this paper, we address the problem of enabling \ac{isac} in realistic \ac{mmwave}
communication systems, by reusing existing communication traffic for sensing as much as possible and thus introducing only a minimal amount of additional overhead. To this end,
we propose SPARCS, the first \ac{mmwave} \ac{isac} system that reconstructs human \ac{md} signatures from \textit{irregular and sparse} \ac{cir} samples obtained from realistic traffic patterns. 
The main insight of SPARCS is to leverage the intrinsic sparsity of the reflections in the \ac{mmwave} channel to pose the \ac{md} reconstruction as a sparse recovery problem. Indeed, \ac{mmwave} \ac{cir} estimation can naturally separate signal propagation paths with $<10$~cm resolution, leading to a sparse multi-path environment and consequently a sparse \ac{cir} in the Doppler domain.
This allows obtaining highly accurate \ac{md} signatures from only a small, randomly distributed fraction of the \ac{cir} samples that are currently needed by existing \ac{isac} methods. 
To do so, SPARCS first performs \ac{cir} resampling to construct a regular grid of \ac{cir} samples with missing vales due to the irregularity of the sampling process in time. Next, a sparse reconstruction method is used to obtain the \ac{md} spectrum, decoupling different propagation paths to leverage their sparsity property.
Lastly, whenever communication traffic is absent or insufficient for the \ac{md} extraction, SPARCS supports a dynamic injection of very short \ac{cir} estimation fields into the (idle) channel. Given its sparse recovery capabilities, only a small number of additional \ac{cir} sensing units are needed to retrieve the \ac{md}, thus entailing a negligible overhead to the communication rate. 

SPARCS is compatible with any \ac{mmwave} system that supports transmit beamforming for directional communication and 
\ac{cir} estimation. This is the case, for example, for IEEE~802.11ay \acp{wlan} at $60$~GHz, which provide in-packet \ac{cir} estimation for beam tracking purposes, and for $3$GPP $5$G-NR, where base stations can send frequent downlink \ac{csirs} to estimate the channel using different \acp{bp}. 

To evaluate SPARCS' performance, we implement it on a $60$~GHz IEEE~802.11ay \ac{sdr} experimentation platform. We then test it on sparse and irregular \ac{cir} samples derived from standard-compliant traces, both for synthetic traffic and traffic patterns obtained from datasets of operational real-world WiFi \acp{ap} deployments \cite{pdx-vwave-20090704}.
To assess the quality of the reconstructed \ac{md} signatures, we use them as input for a typical downstream task such as \ac{har}, which classifies human movement detected by the captured \ac{md} into different possible activities.
The main contributions of our work are summarized next.
\begin{enumerate}
    \item We propose SPARCS, an \ac{isac} method for \ac{mmwave} systems that can reconstruct high-quality \ac{md} signatures of human movement from irregular and sparse \ac{cir} estimation samples. SPARCS reuses training fields appended to communication packets as sensing units, and injects additional sensing units if necessary, adapting to the underlying communication traffic and minimizing the sensing overhead. 
    \item We provide an original formulation of the \ac{md} extraction in communication systems as a sparse recovery problem, leveraging the intrinsic high distance resolution and sparsity properties of the \ac{mmwave} channel. As a side effect, this also improves the quality of the resulting spectrograms, making them more robust to noise and interference.
    \item We design and validate an algorithm to perform the injection of additional sensing units when communication traffic is insufficient. The process is dynamic, requires no knowledge about future packet transmissions, and incurs minimal overall overhead.
    \item We evaluate SPARCS by implementing it on an IEEE~802.11ay-compliant $60$~GHz \ac{sdr} platform and testing it on \ac{cir} measurements collected with realistic WiFi traffic patterns. For the common \ac{har} task, the \ac{md} signatures reconstructed by SPARCS achieve better F$1$ scores than existing methods, while reducing sensing overhead by a factor of $7$.
\end{enumerate}
The paper is organized as follows. In \secref{sec:prel} we discuss the necessary preliminaries for \ac{mmwave} human \ac{md} sensing using \ac{cir}.
SPARCS is introduced and explained in detail in \secref{sec:method}, describing the sparse recovery problem formulation and the involved processing steps.
In \secref{sec:implementation} we discuss the implementation of SPARCS on an \ac{sdr} platform, and \secref{sec:results} provides an evaluation of the system on real measurement traces. We summarize the related work in \secref{sec:rel} and give concluding remarks in \secref{sec:conclusion}.

\section{Primer on mmWave sensing} \label{sec:prel}
In this section we give a brief description of the \ac{cir} model for \ac{mmwave} communication systems that we use for sensing
. We then describe a baseline approach that allows tracking the movement of people in the environment and extract their \ac{md} signatures using \textit{regularly sampled} \ac{cir} information. This forms the basis of our SPARCS design, which 
entirely \textit{eliminates} the requirement of fixed \ac{ifs} and enables ultra low-overhead \ac{isac}.

\subsection{Sensing in \ac{mmwave} systems}\label{sec:sensing-ay} 

Capturing the movement features of humans in the environment requires an analysis of the reflections of the transmitted signal from their bodies, which is usually carried out applying signal processing techniques to the \ac{cir}. Due to the high path loss occurring at \ac{mmwave} frequencies, directional communication is employed by means of transmitter and receiver beamforming, typically using phased antenna arrays. The transmitter and the receiver use suitable \ac{bp} configurations of their antenna arrays to maximize the signal strength 
\cite{giordani2018tutorial, 802.11ad, 802.11ay}.
To successfully sense with a \ac{mmwave} system, \textit{at least} one of the \acp{bp} has to illuminate the subjects of interest, 
as only in this case the reflected signal carries detectable information about the movement signature. To this end, we consider a setup where an \ac{ap} transmits packets and is able to collect the reflections of \textit{its own} signal, after being reflected by objects (including humans). This reflection is collected by the receive array of the \ac{ap} itself using a quasi-omnidirectional \ac{bp}. This requires full-duplex capabilities, as is common in \ac{isac} scenarios \cite{kumari2017ieee}, which in the simplest form can be achieved with a \ac{mimo} system in a mixed configuration with one RF chain as transmitter and another as receiver. The \ac{cir} estimation fields used for sensing, which we denote by \textit{sensing units}, can either be piggybacked by appending them as a trailer to the \ac{phy} communication packets or transmitted independently (\textit{injected}).
\ac{mmwave} standards implement beam training mechanisms that help to establish a communication link by testing different \ac{bp} combinations and then selecting the best one. Such functionality is supported by all \ac{mmwave} standards. For example, $5$G-NR \cite{giordani2018tutorial}, use \ac{ssb} and \ac{csirs} for beam management, while \ac{wlan} systems adopting the IEEE~802.11ad/ay standards \cite{802.11ad, 802.11ay} use channel estimation and training fields (CEF and TRN, respectively) to obtain accurate \ac{cir} information. Our framework to extract sensing information from \ac{cir} measurements can be applied regardless of the specifics of the standards.  


\subsection{\ac{mmwave} CIR model} \label{sec:cir-model}

Due to the large transmission bandwidth of \ac{mmwave} systems, 
channel measurements 
contain fine-grained information about the environment \cite{zhang2020mmeye, pegoraro2021rapid, liu2021integrated}. Depending on the communication system we consider, sensing could be performed using the $5$G-NR \ac{ofdm} \ac{csi}, which contains the channel gains for each \ac{ofdm} subcarrier, or the IEEE~802.11ad/ay \ac{sc} \ac{cir}. Both communication schemes are suitable for human sensing: \textit{(i)} in $5$G-NR, the base stations can send frequent downlink \ac{csirs} to estimate the channel using different \acp{bp}, while \textit{(ii)} in IEEE~802.11ay in-packet beam tracking is enabled, so that specific fields called training fields (TRN), each using a different \ac{bp}, can be appended to communication packets.
In the following, we focus on \ac{sc} \ac{cir}, and show how to extract the \ac{md} effect of human movement. However, previous works have demonstrated that similar processing can be performed with \ac{ofdm} \ac{csi} \cite{meneghello2021environment, liu2021integrated}, and SPARCS is general enough to be applied in both cases. 

\rev{We consider a multipath propagation environment with a time-varying number of reflectors, $P(t)$. These cause physical signal propagation paths that can be separated in the \ac{cir} according to a finite \textit{ranging resolution}, i.e., the capability of the system to resolve the distance of the reflectors causing different signal paths. This is given by $\Delta d = c/2B$, where $B$ is the transmitted signal bandwidth and $c$ the speed of light. Thus, the \ac{cir} contains the complex channel gains for a discrete grid of possible signal paths (or \textit{distance bins}), with indices $\ell= 0,\dots, L-1$. Each path is associated with a specific distance from the \ac{ap}, according to the relation $d_{\ell}=c\tau_{\ell}/2$, with $\tau_{\ell}$ being the delay associated with path $\ell$.} Moreover, the \ac{cir} depends on the specific \ac{bp} used during the transmission, denoted by $b = 0, \dots, N_{\rm BP}-1$. For carrier frequency $f_o$, the \ac{cir} along $\ell$, using \ac{bp} $b$ at time $t$ is
\begin{equation}\label{eq:cir}
    h_{\ell, b}(t) = \sum_{p=1}^{P_{\ell}(t)} a_{\ell, b}^p(t) \exp\left\{-j2\pi \frac{2f_o}{c} \left[d_{\ell} + \int_0^{t} v_{\ell}^p(x) dx\right]\right\}.
\end{equation}
In \eq{eq:cir}, $P_{\ell}(t)$ is the number of physical reflectors whose contributions overlap in the $\ell$-th \ac{cir} path, as their distances are within $d_{\ell} \pm \Delta d /2$, while $v_{\ell}^p$ is the radial velocity\footnote{By convention, $v_{\ell}^p$ has a positive sign when the reflector moves away from the \ac{ap}.} of reflector $p$. The quantity $a_{\ell, b}^p(t)$ is the complex gain due to the joint effect of the transmitter \ac{bp}, the object reflectivity and the signal attenuation. 

\subsection{micro-Doppler extraction} \label{sec:md-extr} 
The extraction of the \ac{md} spectrum from multiple, concurrently moving subjects requires tracking the position of each person in the physical space, in order to separate their individual contributions to the \ac{cir}. Then, a spectral analysis over different \ac{cir} samples yields the desired \ac{md} signature \cite{pegoraro2021rapid, korany2020multiple}. 

\subsubsection{People tracking} \label{sec:track}

\rev{People tracking is performed by extracting measurements of each person's distance and angular position with respect to the \ac{ap} across time, as detailed, e.g., in \cite{pegoraro2021rapid}. This process consists of \textit{(i)} removing the background contribution to the \ac{cir} by subtracting the average \ac{cir} across a suitable time interval, \textit{(ii)} selecting the locally strongest reflection paths in the \ac{cir} (peaks) and obtaining the corresponding distance $d_{\ell}$, \textit{(iii)} computing the \ac{aoa}, $\theta$, of the reflection from the correlation between the different \acp{bp} gains and the strengths of the \acp{cir} across the whole angular \ac{fov} \cite{Lacruz_MOBISYS2020}. The approach in \textit{(iii)} requires the \ac{bp} shapes to be estimated in advance, and is based on the intuition that different \acp{bp} illuminate different possible reflectors in the environment, depending on their position. Therefore, we can expect to receive the reflected signal from a certain subject only when a \ac{bp} is pointing in his/her direction.}
The resulting set of distances and angles represent candidate positions of humans in the environment. A multi-target tracking method such as a \ac{jpdaf} \cite{shalom2009probabilistic} allows smoothing the trajectories of the subjects by rejecting noise and clutter. 

The tracking phase provides an estimate of the position of each subject, at every time instant $t$, which we denote by $[\hat{d}(t), \hat{\theta}(t)]$
using symbol $\hat{}$ to differentiate between our estimates and the true position. Due to step $(ii)$ above, which selects the local peaks in the received power, typically only the reflections from the torso can be tracked when a person moves in the environment. In contrast, the head and the limbs cause much weaker reflections \cite{vandersmissen2018indoor}, whose contributions can only be detected in the \ac{md} spectrum. 
Nevertheless, tracking the spatial location of the torso is crucial for the \ac{md} extraction as it allows separating the contributions of the multiple subjects in the environment. 
Once $\hat{d}(t)$ and $\hat{\theta}(t)$ are determined, they are used to select the path $\ell^*$ and the \ac{bp} $b^*$ that correspond to the distance and angular position of the subject, respectively. Then, the \ac{cir} waveform that contains the \ac{md} effect of the person's movement is $h_{\ell^*, b^*}(t)$. \rev{Besides enabling the separation of multiple subjects, this operation makes mmWave sensing systems much more robust to changes in the environment than sub-$6$~GHz systems \cite{meng2020gait}. While the latter are heavily affected by second-order reflections on walls and objects, mmWave sensing mostly relies on the line-of-sight path between the person and the transmitter/receiver, with little contribution from the external environment.}

\subsubsection{micro-Doppler spectrum} \label{sec:md}
Human movement causes a small-scale Doppler effect on the reflected signal due to the different body parts, which possess different velocities and follow different trajectories \cite{chen2006micro}. This is referred to as \ac{md} effect and causes a measurable frequency modulation on the reflection. High frequency signals such as \ac{mmwave} communications are particularly affected by the \ac{md} modulation due to their small wavelengths. Various techniques from time-frequency analysis can be applied to analyze the \ac{md}, obtaining \textit{spectrograms} showing the time evolution of the signal energy contained in the different frequency bands of interest. The most popular and computationally efficient of such methods is the \ac{stft} of $h_{\ell, b}(t)$ \cite{boashash2015time}, which consists in applying a windowed \ac{ft} on partially overlapping portions of the \ac{cir}.
As shown in \cite{pegoraro2021rapid, meneghello2021environment, korany2020multiple}, \ac{stft} processing needs to be applied to a window of $W$ subsequent estimates of the \ac{cir} with a \textit{fixed} \ac{cir} sampling interval of $T_c$ seconds, provided that the time spanned by the window is short enough to consider the movement velocity of the reflectors \textit{constant} for its whole duration. Note that this operation allows detecting and separating the velocities of the $P_{\ell}(t)$ reflectors, whose contributions overlap in path $\ell$ when considering a single estimate of the \ac{cir}.
The choice of $T_c$ impacts the frequency resolution of the \ac{stft}, $\Delta f^{d} = 1/(WT_c)$, and its maximum measurable frequency, $f^d_{\rm max} = 1/(2T_c)$. Using the relationship between the Doppler frequency and the corresponding velocity, one can obtain the velocity resolution and the maximum observable velocity as $\Delta v = c/(2f_o  W T_c)$ and $v_{\rm max} = c/(4f_o T_c)$. To fully capture the range of velocities of interest for human movement, the typical approach is to select $T_c$ such that $v_{\rm max}$ is sufficiently high that is covers the velocities that can occur in the human activities of interest, which may vary depending on the application  \cite{vandersmissen2018indoor, pegoraro2021rapid, singh2019radhar}.

Previous work assumes that the constraint of a fixed $T_c$ is met, which does not hold in realistic communication scenarios, where packet transmissions are scheduled according to the needs of the communication protocols rather than sensing accuracy. Traffic patterns are typically bursty and irregular and thus cannot be used by existing methods for human sensing. Instead, dedicated time slots need to be reserved for the transmission of sensing units, which is incompatible with the random access CSMA/CA MAC commonly used in IEEE 802.11. 
Conversely, SPARCS is the first approach that does not require any specific pattern in the transmission of the sensing units, enabling true \ac{isac} by exploiting communication packets for sensing whenever possible, and introducing minimal additional overhead when necessary.

\section{SPARCS methodology} 
\label{sec:method}
\begin{figure}
     \centering
     \includegraphics[width=8.4cm]{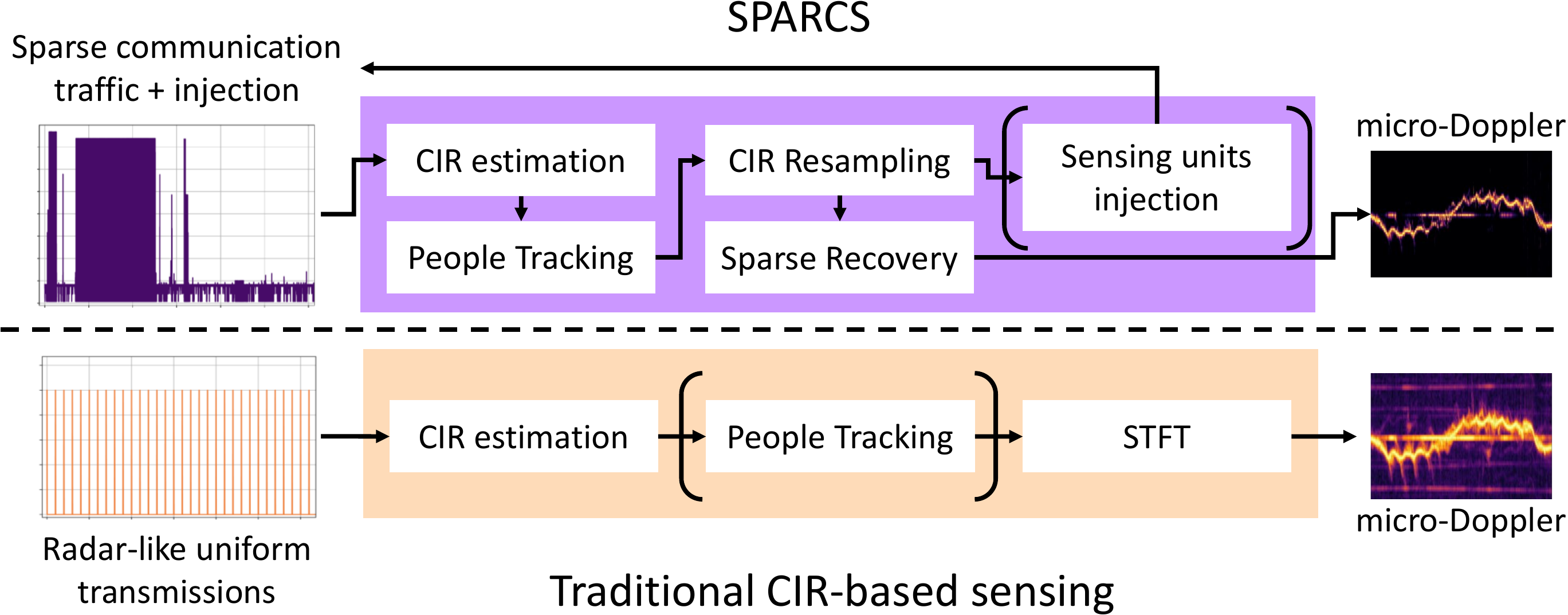}
     \caption{Comparison between the traditional \ac{cir}-based human sensing and SPARCS.}
     \label{fig:workflow}
\end{figure}
We now present the SPARCS algorithm to recover the \ac{md} spectrum from irregular and sparse \ac{cir} sampling patterns. 
The processing steps of SPARCS compared to traditional \ac{cir}-based sensing methods are shown in \fig{fig:workflow}.

\textbf{$(1)$ \ac{cir} resampling:} after \ac{cir} estimation and people tracking, for which we adopt the standard \ac{jpdaf} technique \cite{shalom2009probabilistic}, we apply a resampling strategy to approximate the irregularly spaced \ac{cir} values with a regular sequence whose sampling interval is chosen according to the desired \ac{md} resolution (\secref{sec:cir-resample}). Due to irregularity of the original sampling process, the approximated regular sequence may contain missing values that need to be \textit{filled} in the subsequent processing steps.

\textbf{$(2)$ Sparse \ac{md} recovery:} we formulate the recovery of the \ac{md} spectrum from the incomplete \ac{cir} measurements as a sparse recovery problem. \rev{For this, we leverage two key aspects. On the one hand, the intrinsic sparsity of the \ac{mmwave} channel leads to a small number of signal reflections from the human body that carry information about different body parts. On the other hand, the high distance resolution of \ac{mmwave} systems makes the reflections from the different body parts separable. The combined effect of these two properties is that the resulting \ac{cir} is highly sparse in the Doppler frequency domain, as detailed in \secref{sec:sparse-rec}}
We then solve the sparse recovery problem using the \ac{iht} algorithm for each \ac{cir} path (\secref{sec:single-path}), and aggregate the results to obtain the final \ac{md} spectrum (\secref{sec:multipath-acc}).

\textbf{$(3)$ Sensing unit injection:} when communication traffic is absent or too scarce to obtain an accurate reconstruction, our system can inject short sensing units into the (idle) channel to overcome the problem, as described in \secref{sec:injection}. Thanks to the sparse reconstruction of point $(2)$, the amount of units that need to be injected is minimal and can be tuned to trade off between overhead and sensing accuracy.

\subsection{CIR resampling}
\label{sec:cir-resample}
Our system samples the \ac{cir} at time instants $t_i$, which coincide with the reception of the reflections from the $i$-th transmitted 
packet.
To reconstruct the \ac{md} spectrum from \ac{cir} samples which are randomly distributed in the time domain, we first resample the \ac{cir} to obtain regularly spaced samples with a fixed granularity $T_c$, where possible.
To do so, we resort to the \textit{slotted resampling} technique, which allows approximating a sequence of randomly spaced samples into a regular grid with \textit{missing values} \cite{babu2010spectral}. 
We consider $N_s$ consecutive samples obtained at the time instants $t_0, t_1, \dots, t_{N_s}$ and denote by $0, T_c, 2T_c, \dots, (K-1)T_c$ the regular grid with step size $T_c$.
Slotted resampling constructs a new \ac{cir} sample sequence $h_{\ell, b}(kT_c)$ where the \ac{cir} values are obtained from the original sequence $h_{\ell, b}(t_i)$ as follows.
Time \textit{bins} (or intervals) of length $T_c$ are centered on each time instant of the regular grid, i.e., bin $k$ is $\beta_k = [kT_c - T_c/2, kT_c + T_c/2)$, with center $kT_c$.
Then, the value of the \ac{cir} corresponding to the $k$-th grid value is either \textit{(i)} selected among the values of the original sequence whose sampling times fall inside bin $k$, taking the one whose sampling time is the closest to the bin center, or \textit{(ii)} considered as a missing value if no samples of the original sequence fall inside bin $k$.
Specifically,
\begin{equation}
    h_{\ell, b}(kT_c) =  \left\{
\begin{array}{cc}
0 &\mbox{ if } \{t_i | t_i \in \beta_k\} = \emptyset,\\
  h_{\ell, b}(t_k) &   \mbox{otherwise},
\end{array}\right.
\end{equation}
where the $0$ values represent missing samples and
\begin{equation}
    t_k = \argmin_{\tau \in \{t_i | t_i \in \beta_k\}} |kT_c - \tau|.
\end{equation}

The resulting, regularly spaced sequence of \ac{cir} samples is used to reconstruct the \ac{md} spectrum of the subject. However, due to the missing samples which are set to $0$, a plain application of the \ac{stft} (as described in \secref{sec:md-extr}) would lead to a corrupted spectrum. In the next section we detail our solution to this problem, which is based on sparse recovery techniques.

\subsection{Sparse \ac{md} recovery problem formulation}\label{sec:sparse-rec}

Several methods exist to tackle the problem of computing the power spectrum of non-uniformly sampled signals \cite{babu2010spectral}. Our approach belongs to the category of sparsity-based approaches, in which the sparsity of the signal in the frequency domain is leveraged to drastically reduce the number of measurements needed for an accurate reconstruction of the spectrum.
We select windows of length $W$ samples (window size) every $\delta$ samples from the sequence $h_{\ell, b}(kT_c)$, choosing $\delta=W/2$. Due to the slotted resampling process, each window may contain missing samples. We denote by $\mathcal{U}_m$ the set of indices of the available 
samples contained in the $m$-th window.
Then, we define vector $\mathbf{h}_{\ell, b}(m) \in \mathbb{C}^{|\mathcal{U}_m|}$, containing the available \ac{cir} samples in the $m$-th window, and vector $\tilde{\mathbf{h}}_{\ell, b}(m) \in \mathbb{C}^{W}$, representing the complete $m$-th \ac{cir} window, which is only partially known due to the missing samples. 
We also denote by $\mathbf{F}_{\rm inv}$ the inverse Fourier matrix, whose element in position $(g, l)$ is given by $\left( \mathbf{F}_{\rm inv}\right)_{gl} = (1/\sqrt{W})\exp \left(j 2\pi gl/W\right), \,\, g, l = 0, \dots, W - 1$
while $\mathbf{U}_m=\left[ \mathbf{u}^T_{i}\right], \forall i \in \mathcal{U}_m$ is the matrix that selects the rows of $\mathbf{F}_{\rm inv}$ whose indices are in $\mathcal{U}_m$. $\mathbf{u}_{i}$ is the vector of all zeros but the $i$-th component, which equals $1$. 

The following relation holds between the incomplete \ac{cir} window, $\mathbf{h}_{\ell, b}(m)$, and the \ac{ft} of the full \ac{cir} window, $\mathbf{H}_{\ell, b}(m) \in \mathbb{C}^{W}$, which we aim to recover in order to compute the \ac{md} spectrum,
\begin{equation}\label{eq:cs-model}
    \mathbf{h}_{\ell, b}(m) = \mathbf{U}_m \tilde{\mathbf{h}}_{\ell, b}(m)=  \mathbf{U}_m  \mathbf{F}_{\rm inv}\mathbf{H}_{\ell, b}(m) = \mathbf{\Psi}_m \mathbf{H}_{\ell, b}(m),
\end{equation}
where in the last step we use matrix $\mathbf{\Psi}_m = \mathbf{U}_m  \mathbf{F}_{\rm inv}$ as a shorthand notation.
Given \eq{eq:cs-model}, our aim is to recover $\mathbf{H}_{\ell, b}(m)$ from the incomplete measurement vector $\mathbf{h}_{\ell, b}(m)$, which is a typical sparse recovery or compressed sensing problem \cite{eldar2012compressed}. In this framework, it has been proven that recovering the \ac{ft} of the desired signal is possible if the latter is sparse in the frequency domain, i.e., the \ac{ft} only contains a low fraction of non-zero elements. 
To verify that this sparsity assumption holds in our case, we rewrite \eq{eq:cir} after the resampling and windowing operations, so that the $i$-th sample of the complete $m$-th window is given by
\begin{equation}\label{eq:cir-discrete}
    \left[\tilde{\mathbf{h}}_{\ell, b}(m)\right]_{i} = \sum_{p=1}^{P_{\ell}(m)}a_{\ell, b}^p(m) \exp\left\{-j4\pi \frac{f_o}{c} \left[d_{\ell}^p + (m\delta+i) T_c v_{\ell, m}^p \right]\right\},
\end{equation}
where $v_{\ell, m}^p$ is the radial velocity of the $p$-th reflector in path $\ell$ during window $m$, and $d_{\ell}^p$ its distance from the \ac{ap}. 
Here, we use the assumption from \secref{sec:md-extr} that the velocity of each reflector can be considered constant during a window. In addition, we also consider that the reflective coefficients and the number of reflectors are constant. This is reasonable for the considered setup, where the reflectors are parts of the human body, which typically move slowly compared to the duration of a window $WT_c$ (see also \secref{sec:implementation}).

\begin{figure}
     \centering
     \includegraphics[width=8.5cm]{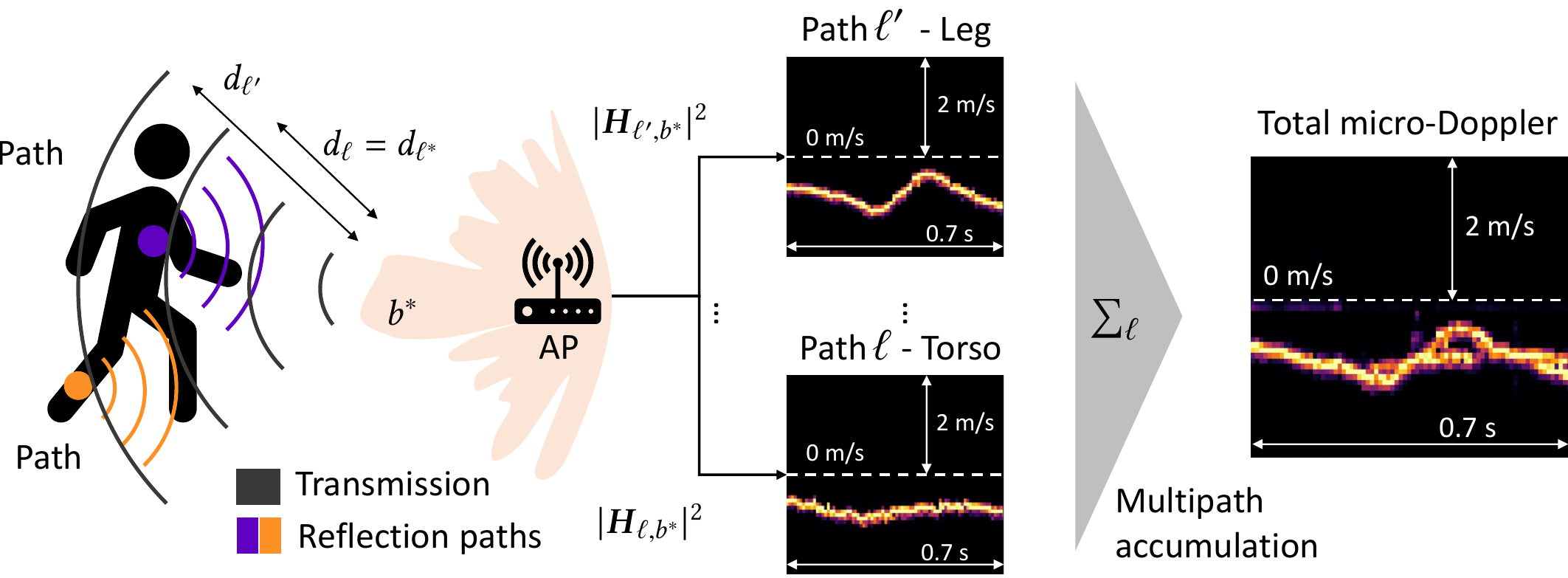}
     \caption{Visual representation of the \ac{md} spectrum computed using SPARCS on $2$ different \ac{cir} paths, one containing the reflection from a person's torso, the other capturing the \ac{md} signature of the leg. The total \ac{md} is obtained summing together these contributions.}
     \label{fig:separation}
\end{figure}

From \eq{eq:cir-discrete}, one can see that as long as $P_{\ell}(m) \ll W$, the \ac{ft} of $\tilde{\mathbf{h}}_{\ell, b}(m)$ is indeed sparse, as it is composed of $P_{\ell}(m)$ spectral lines located at frequencies $2f_o v_{\ell, m}^p / c$.
Given the excellent distance resolution due to the high bandwidth of \ac{mmwave} systems and the \textit{intrinsic} sparsity and directionality of the channel, the different parts of the subject's body tend to contribute to the \ac{md} spectrum in different \ac{cir} paths as shown in \fig{fig:separation}. Therefore, $P_{\ell}(m)$ is generally close, if not equal, to $1$. Sometimes the number of reflectors in a single path can be larger than $1$, due to different body parts being closer than the distance resolution of the system, but this number is still much lower than $W$.
\rev{This even holds for multiple subjects. Assume that two subjects with labels $1$ and $2$ are present in the monitored physical space, and denote by $(\ell_1, b_1)$ and $(\ell_2, b_2)$ their \ac{cir} path-\ac{bp} pairs. According to \eq{eq:cir-discrete}, the sparsity assumption must hold for each pair \textit{independently}, and this is verified as long as the subjects occupy different spatial positions. Specifically, \textit{(i)} if $\ell_1 \neq \ell_2$ the \acp{cir} along \acp{bp} $b_1$ and $b_2$ are the combination of $P_{\ell_1}(m)\ll W$ and $P_{\ell_2}(m)\ll W$ complex exponentials each, and \textit{(ii)} if $\ell_1 = \ell_2$, but $b_1 \neq b_2$, the attenuation coefficient of $b_1$ will mostly remove the reflection from subject $2$ in $\tilde{\mathbf{h}}_{\ell_1, b_1}$ and vice versa, making the contributions from the subjects separable. The contributions from different subjects overlap only if they occupy the same \ac{cir} path \textit{and} share the same \ac{bp}, which is very unlikely to occur in real cases due to the high distance ($\sim 8$~cm) and angular (as low as $ 2^{\circ}$) resolutions of the \ac{mmwave} \ac{cir} \cite{pegoraro2021rapid}. 
Therefore, the sparsity assumption in SPARCS still holds even if multiple subjects are present in the environment.}
Due to this, we can assume that $P_{\ell}(m) \ll W$ holds, and that sparse recovery techniques can be used to recover $\mathbf{H}_{\ell, b}(m)$, as detailed in the next section. 

\subsection{Single-path sparse recovery}
\label{sec:single-path}

\begin{algorithm}[t!]
\footnotesize
	\caption{Single path sparse recovery.}
	\label{alg:sparse-recovery}
	\begin{algorithmic}[1]
		\REQUIRE $\mathbf{h}_{\ell, b^*}(m)$, $\eta, n_{\rm max}, \Omega, \xi$.
		\ENSURE $\mathbf{H}_{\ell, b^*}(m)$.
		\STATE Collect the set of available samples indices $\mathcal{U}_m$.
		\STATE Build matrices $\mathbf{U}_m=\left[ \mathbf{u}^T_{i}\right], \forall i \in \mathcal{U}_m$ and $\mathbf{\mathcal{F}}_{\rm inv}$
		\STATE Compute $\mathbf{\Psi}_m = \mathbf{U}_m  \mathbf{\mathcal{F}}_{\rm inv}$.
		\STATE Set $\hat{\mathbf{H}}^{(0)} = \mathbf{0}$, $ n = 0$, $\gamma^{(0)}$ to any value $> \xi$.
		\WHILE{$ n < n_{\rm max}$ or $\gamma^{(n)} > \xi$}
		\STATE $\hat{\mathbf{H}}^{(n+1)} \leftarrow $ \eq{eq:iht}
		\STATE $\gamma^{(n + 1)} \leftarrow ||\hat{\mathbf{H}}^{(n+1)} - \hat{\mathbf{H}}^{(n)}||_2$
		\STATE $n \leftarrow n+1$
		\ENDWHILE
		\RETURN $\hat{\mathbf{H}}^{(n)}$
	\end{algorithmic}
\end{algorithm}
Given the model from \eq{eq:cs-model}, the reconstruction of the \ac{cir} \ac{ft} along each path can be posed as a sparse recovery problem. Specifically, we seek a vector $\mathbf{H}_{\ell, b}(m)$ which is a solution to \eq{eq:cs-model} while being as sparse as possible, coherent with the above discussion.
Considering the \ac{bp} $b^*$ pointing in the direction of the target, the desired \ac{ft} of $\tilde{\mathbf{h}}_{\ell, b^*}(m)$ is the solution of the optimization problem
\begin{equation}\label{eq:sp-opt}
    \mathbf{H}_{\ell, b^*}(m)= \argmin_{\mathbf{H}} ||\mathbf{H}||_0 \quad \mbox{   s.t.   } ||\mathbf{h}_{\ell, b^*}(m) - \mathbf{\Psi}_m\mathbf{H}||_2 \leq \varepsilon,
\end{equation}
where $||\cdot||_0$ denotes the $\ell_0$-norm of a vector, i.e., the number of its non-zero components. The constant $\varepsilon > 0$ can be estimated from the noise in the \ac{cir}, using a training dataset.

An approximate local solution to \eq{eq:sp-opt} can be found using fast greedy algorithms \cite{eldar2012compressed}. We adopt the \ac{iht}, which solves
\begin{equation}\label{eq:local-opt}
    \mathbf{H}_{\ell, b^*}(m)= \argmin_{\mathbf{H}} ||\mathbf{h}_{\ell, b^*}(m) - \mathbf{\Psi}_m\mathbf{H}||_2^2 \quad \mbox{   s.t.   } ||\mathbf{H}||_0  \leq \Omega,
\end{equation}
where $\Omega$ is a pre-defined sparsity level parameter.
The algorithm involves an iterative gradient descent step on the quadratic term in \eq{eq:local-opt}, followed by a thresholding operation: 
\begin{equation}\label{eq:iht}
    \hat{\mathbf{H}}^{(n+1)} \leftarrow \mathcal{T}_{\Omega}\left[ \hat{\mathbf{H}}^{(n)} + \eta \mathbf{\Psi}_m^T\left( \mathbf{h}_{\ell, b^*}(m) - \mathbf{\Psi}_m\hat{\mathbf{H}}^{(n)} \right)\right],
\end{equation}
where $n$ is the iteration index and $\mathcal{T}_{\Omega}$ is the hard-thresholding operator, which sets to $0$ all the components of the argument vector except the $\Omega$ largest ones in terms of the Euclidean norm. $\eta$ is a learning rate parameter which can be tuned to improve the convergence properties.
The iterative process is stopped whenever $||\hat{\mathbf{H}}^{(n+1)} - \hat{\mathbf{H}}^{(n)}||_2 < \xi$ or when a maximum number of iterations, $n_{\rm max}$, is reached.
In SPARCS, $\Omega$ is a key parameter, which is strictly related to the number of reflectors $P_{\ell}(m)$: as \ac{iht} reconstructs a vector which has at most $\Omega$ non-zero elements, $\Omega$ is an upper bound for $P_{\ell}(m)$, and it can be thought of as the maximum number of reflectors per path that we allow reconstructing.
$\Omega$ can be tuned in order to obtain better \ac{md} reconstruction (see \secref{sec:params-sens}).
The sparse recovery algorithm is summarized in \alg{alg:sparse-recovery}.
\rev{According to the compressive sensing theory \cite{foucart2013mathematical}, 
the reconstruction performance of \ac{iht} (and in general of any recovery algorithm) degrades as the number of available measurements, $|\mathcal{U}_m|$, decreases. Theoretical results show that the minimum number of measurements needed to reconstruct $\mathbf{H}_{\ell, b^*}(m)$ is  $\mathcal{O}(\Omega \log(W/\Omega))$ \cite{foucart2013mathematical}, although the exact number has to be estimated empirically as it also depends on the level of noise present in the signal. In \secref{sec:results}, we show that SPARCS can achieve excellent \ac{md} reconstruction with as few as $W/8$ measurements per window, thanks to the high sparsity of the \ac{mmwave} \ac{cir}.}

\subsection{Multi-path aggregation}\label{sec:multipath-acc}

The moving body of a person causes several reflections that affect more than one \ac{cir} path, as discussed in \secref{sec:sparse-rec}. Using the procedure described in the previous sections, SPARCS is able to retrieve the contribution of each path $\mathbf{H}_{\ell, b^*}(m)$ to the \ac{md}. Since the different body parts contribute to the \ac{md} in different paths, to fully capture human movement we need to combine the information from the different paths. 
Denote by $Q$ the number of distance bins we aggregate to obtain the \ac{md} spectrum. For convenience we assume $Q$ to be an odd integer, as this makes the following processing steps symmetric with respect to a central \ac{cir} path (corresponding to the torso), but the same steps can be applied for $Q$ being even. We aggregate the spectra obtained from the path caused by the torso, $\ell^*$, with the $\lfloor Q/2 \rfloor$ distance bins preceding $\ell^*$ and the $\lfloor Q/2 \rfloor$ subsequent distance bins, as they may contain the contributions of the other body parts.
The expression of the total \ac{md} spectrum is
\begin{equation}\label{eq:sparse-paths-sum}
   \mathbf{D}(m) =\sum_{\ell=\ell^* - \lfloor Q/2 \rfloor }^{\ell^* + \lfloor Q/2 \rfloor} \left|\mathbf{H}_{\ell, b^*}(m)\right|^2,
\end{equation}
where the squared magnitude is applied element-wise. In addition, we apply normalization of the spectra in the range $[0, 1]$ by computing $\mathbf{D}(m) \leftarrow \frac{\mathbf{D}(m) - \min_i  \mathrm{D}_i(m)} {\max_i \mathrm{D}_i(m) - \min_i \mathrm{D}_i(m)}$.
Note that \eq{eq:sparse-paths-sum} entails solving $Q$ optimization problems of the form in \eq{eq:local-opt}, however, the $Q$ problems can be parallelized as they are completely independent. 
Decomposing the full \ac{md} spectrum reconstruction problem into $Q$ subproblems effectively allows applying sparse recovery techniques, which in turn leads to a significant reduction of the number of measurements that are needed.

The value of $Q$ is selected according to physical considerations and validated in practice, as described in \secref{sec:results}.
The \ac{md} vectors from \eq{eq:sparse-paths-sum} can be collected in sequences, one every $\delta$ slots, forming \ac{md} spectrograms of arbitrary length, depending on the specific application that is being performed, e.g., activity recognition, fall detection, gait segmentation, etc. In the following, we refer to the number of \ac{md} vectors considered in such spectrograms as $\Lambda$.

\subsection{Sensing unit injection}
\label{sec:injection}

SPARCS can exploit the sensing units in sparsely distributed communication packets to recover the \ac{md} spectrum of human movement. \rev{However, during communication between the \ac{ap} and one or more terminals it may happen that the \ac{ap} remains silent for longer than the duration of a processing window, $WT_c$, or that the received packets are fewer than the minimum number of measurements required for an accurate \ac{md} reconstruction. In these cases, the sparse recovery algorithm can not recover $\mathbf{H}_{\ell, b}(m)$ as the available sensing units are insufficient.}
To tackle this problem, we allow our system to \textit{inject} sensing units into the channel whenever the number of communication packets is not sufficient for \alg{alg:sparse-recovery} to work. Different from existing \ac{isac} frameworks, 
our sparse recovery approach allows us to introduce a minimal amount of overhead, as the \ac{md} spectrum can be recovered from a number of \ac{cir} samples which is much lower than the full length of the window $W$. Note that for the injection of a sensing unit it is sufficient to transmit the necessary \ac{cir} estimation fields, without any preamble and header as used in conventional packets, since the unit is only received at the \ac{ap} itself and contains a known waveform.

\subsubsection{Basis of the injection algorithm} \label{sec:inj-setup}
In the following, we present the proposed injection procedure assuming that both communication packets and sensing units are transmitted at times that lie on a uniform grid with spacing $T_c$. This simplification is valid due to the fact that the slotted resampling process described in \secref{sec:cir-resample} is used.
Therefore, we can describe the injection process in terms of windows of size $W$, where each value in the window occupies a \textit{slot} which is a multiple of $T_c$. Due to slotted resampling, the slots can be empty if no packet was transmitted sufficiently close to it.

Our approach consists in setting a minimum number of sensing units \textit{per window}, termed $M_{s}$, that allows a sufficiently accurate reconstruction of the \ac{md} signatures. We then transmit additional units whenever the number of reflections of communication packets in the window is not sufficient to meet this minimum requirement. 
The proposed method only requires the knowledge of whether a reflected communication packet is received in the \textit{current} slot, i.e., no information about the future traffic pattern is needed. 


\subsubsection{Algorithm description} \label{sec:inj-algo}
\begin{algorithm}[t!]
\footnotesize
	\caption{Injection of sensing units in window $m$.}
	\label{alg:inj-algo}
	\begin{algorithmic}[1]
		\REQUIRE $M_s$.
		\STATE \textcolor{gray}{\texttt{\# P1 - observation phase}}
		\STATE $N_a(m) \leftarrow$ no. of sensing units received in the first half of the window (either from reflected communications packets or injected).
		\STATE \textcolor{gray}{\texttt{\# P2 - scheduling phase}}
		\STATE $N_w(m) \leftarrow \max(M_s - N_a(m), 0)$.
		\STATE Schedule $\mathcal{S}_m=\left\{s_1, \dots, s_{N_w(m)}\right\}$.
		\STATE \textcolor{gray}{\texttt{\# P3 - transmission phase}}
		\FOR{$q = m W/2, \dots, (m+1)W/2 - 1$}
		\IF{$q \in \mathcal{S}_m$}
		\IF{no reflected comm. packet received}
		\STATE Transmit the sensing unit.
		\STATE $\mathcal{S}_m \leftarrow \mathcal{S}_m \setminus \{q\}$.
		\ELSE
		\STATE Use the sensing unit from the comm. packet
		\STATE $\mathcal{S}_m \leftarrow \mathcal{S}_m \setminus \{q\}$.
		\ENDIF
		\ELSE
		\IF{reflected comm. packet received}
		\STATE Use the sensing unit from the comm. packet
		\STATE $\mathcal{S}_m \leftarrow \mathcal{S}_m \setminus \{\min_{s \in \mathcal{S}_m} s\}$.
		\ENDIF
		\ENDIF
		\ENDFOR
		\end{algorithmic}
\end{algorithm}
The algorithm, summarized in \alg{alg:inj-algo}, operates in three phases, namely \textit{observation} (P$1$), \textit{scheduling} (P$2$) and \textit{transmission} (P$3$). Recall that the \ac{md} extraction described in \secref{sec:sparse-rec} follows a window-based approach, with subsequent windows overlapping by half of their length, as shown in \fig{fig:injection}. 
Consider a time instant between the end of window $m-1$ and the start of window $m+1$. This coincides with the \textit{half} of window $m$, which is between slots $m W/2-1$ and $m W/2$. 
In this time instant we can observe how many reflected communication packets were received in the first half of window $m$, which spans the indices from $(m-1)W/2$ to $m W/2-1$ (P$1$, line $2$ in \alg{alg:inj-algo}). We denote this number as $N_a(m)$. 
The injection algorithm is executed on a half-window basis at the time when window $m-1$ has ended and window $m+1$ has not yet started, as this allows reasoning on the sole current window $m$.
Based on $N_a(m)$, we can compute how many sensing units we would need in the remaining half of window $m$ in order to meet the requirement of at least $M_{s}$ units, which we denote by $N_w(m) = \max(M_s - N_a(m), 0)$. However, the sensing process has no knowledge of when future communication packets will be received, so the best we can do is schedule the transmission of $N_w(m)$ sensing units in the next half-window. The slots in which these packets are scheduled can be selected according to a deterministic rule or a probability distribution. We call $\mathcal{S}_m = \left\{s_1, \dots, s_{N_w(m)}\right\}$ the set of indices of the slots in which we schedule the additional sensing units for the next half-window (P$2$, lines $4$-$5$ in \alg{alg:inj-algo}).
While P$1$ and P$2$ are performed in a single time slot, before the second half of window $m$ starts, P$3$ (lines $7$-$23$ of \alg{alg:inj-algo}) is a dynamic process that spans the whole second half of window $m$. The indices of the slots considered in this part of the algorithm are $q = mW/2, \dots, (m+1)W/2 - 1$. Note that some communication packets, of which we have no knowledge, may be received in this second half-window. The procedure iterates over the slots and in each of them checks if a sensing unit was scheduled for that slot, i.e., if $q \in \mathcal{S}_m$. There are four possible cases:

\begin{figure}
     \centering
     \includegraphics[width=7.5cm]{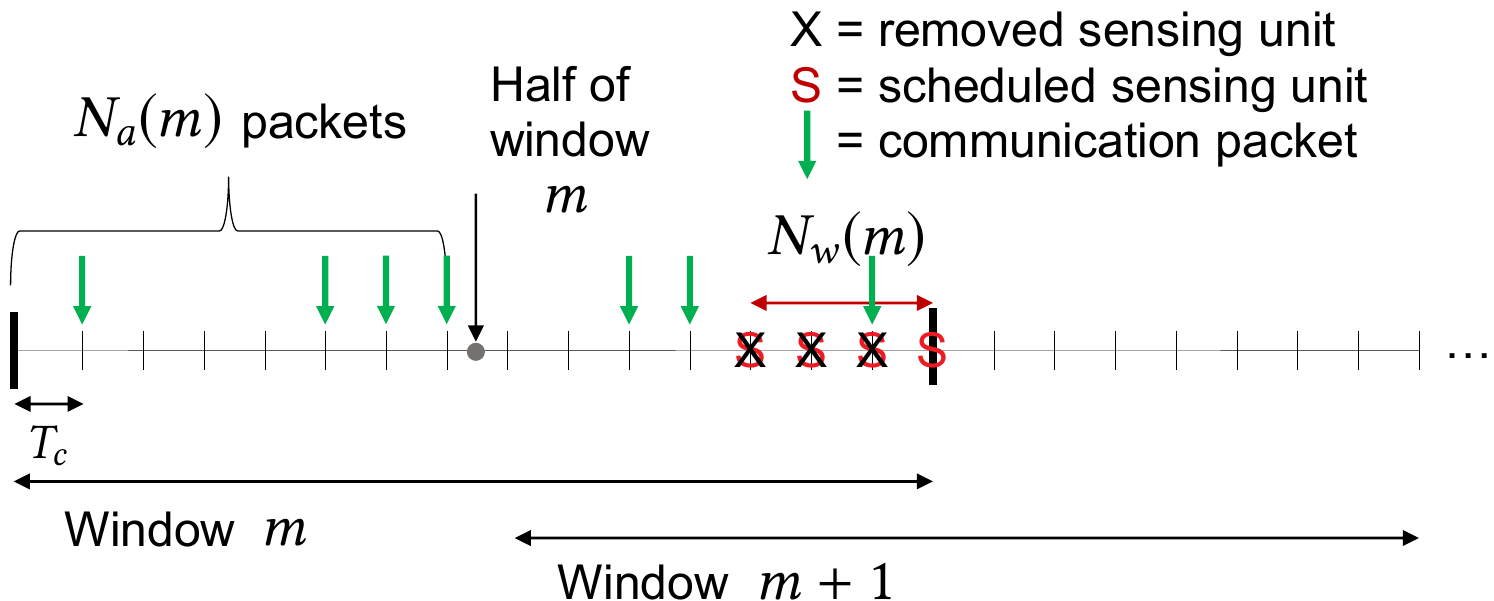}
     \caption{Example injection procedure with $M_s=8, W=16$. $4$ sensing units are scheduled after P$1$ and P$2$. Then, as three reflected communication packets are received, we reuse them and the first three scheduled sensing units are not transmitted. The fourth sensing unit is instead injected in the last slot.}
     \label{fig:injection}
\end{figure}

\noindent $\mathbf{(1)}$ $q \in \mathcal{S}_m$ and no communication packet was received in this slot. In this case we transmit the sensing unit, then remove $q$ from $\mathcal{S}_m$.

\noindent $\mathbf{(2)}$ $q \in \mathcal{S}_m$ and a communication packet (or more) was received in this slot. In this case we reuse the sensing unit in the communication packet and remove $q$ from $\mathcal{S}_m$. 

\noindent $\mathbf{(3)}$ $q \notin \mathcal{S}_m$ and no communication packet was received in this slot. In this case we just move to the next slot without taking action.
    
\noindent $\mathbf{(4)}$ $q \notin \mathcal{S}_m$ and a communication packet (or more) was received in this slot. In this case we reuse the sensing unit in the communication packet, then we remove the next sensing unit from the scheduled ones, i.e., we set $\mathcal{S}_m \leftarrow \mathcal{S}_m \setminus \{\min_{s \in \mathcal{S}_m} s\}$.

Note that, despite operating on a half-window basis, due to the overlap of adjacent windows, our algorithm only poses a constraint on the minimum number of packets sent per \textit{full window}. This means that half a window can be empty as long as enough sensing units are received in the other half.

\subsubsection{Scheduling the sensing units}\label{sec:scheduling}
While the scheduling of the sensing units in P$2$ can be done with any arbitrary policy that guarantees that exactly $N_w(m)$ packets are scheduled in the next half window, we want to maximize the number of sensing units that can be piggybacked on communication packets, rather than using a dedicated transmission.
From P$3$ in \alg{alg:inj-algo}, one can see that scheduling the sensing units towards the end of the half-window leaves more time for possible communication packets to become available and thus be reused instead of injecting a new sensing unit.
Consequently, in SPARCS we schedule the 
sensing units for the second half of window $m$ as a burst of packets spaced by $T_c$, which occupy the last $N_w(m)$ slots of the window.

\section{Implementation}\label{sec:implementation}

In this section we describe the implementation of SPARCS on a \ac{mmwave} \ac{sdr} platform. We base our implementation on the IEEE~802.11ay WiFi protocol, as it operates in the unlicensed $60$~GHz band and supports \ac{cir} estimation for different \acp{bp}.

\noindent \textbf{Testbed.} We use the open-source mm-FLEX experimentation platform from \cite{Lacruz_MOBISYS2020} for a baseline design. It has an FPGA-based baseband processor which can generate, capture and process (custom or standard compliant) frames with up to $1.76$~GHz of bandwidth. The baseband processor is connected to \ac{mmwave} RF front-ends with phased antenna arrays and supports various front-ends to operate in different frequency bands, e.g., at $28$~GHz or $60$~GHz \cite{SIVERSIMA_BOTH}. In the remainder of this paper we use a $60$~GHz RF front-end which simplifies experimentation as this is an unlicensed band, but we note that simply by changing the RF front-end, SPARCS can operate in a different band, e.g., for 5G-NR compatibility. 

\begin{figure}[t!]
    \centering
    \includegraphics[width=0.9\linewidth]{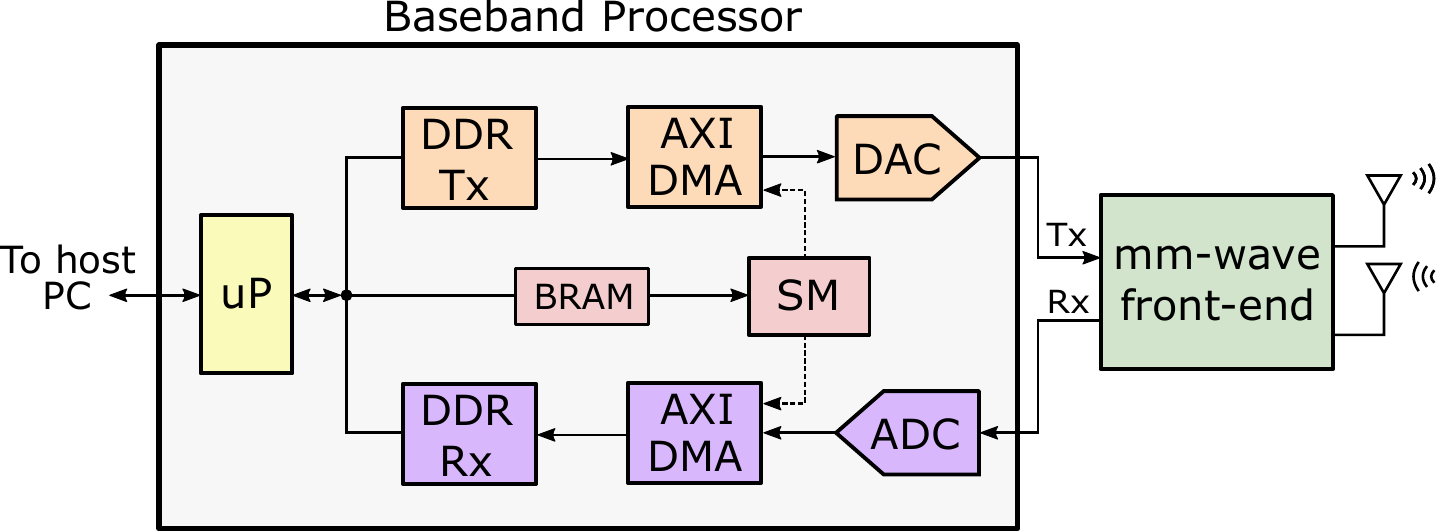}
    \caption{SPARCS implementation block diagram.}
    \label{fig:implementation}
\end{figure}

To implement SPARCS, we augment the functionalities of the testbed to enable full-duplex operation, not only in the baseband processor but also in the RF front-end. A block diagram of the system is shown in \fig{fig:implementation}. In the baseband processor, the operation of the system is controlled by a state machine (SM) which triggers an AXI-DMA that reads the I/Q samples from the DDR-TX memory and feeds them to the DACs. The SM also triggers another AXI-DMA in the receiver datapath that saves the receives samples in the DDR-RX memory, i.e., \textit{both} datapaths are synchronized and no packet detection is required. To support the variable \ac{ifs} extracted from real (or artificially generated) traces, we include a block RAM memory (BRAM) in the FPGA logic that stores the \ac{ifs} that will be used in the experiments. The SM reads these values sequentially, introducing a delay in the system according to the value read from memory. The variable \ac{ifs} functionality can be disabled at runtime to configure a \textit{fixed} \ac{ifs}. 
We remark that since we simultaneously use the up/down conversion stages from the \textit{same} \ac{mmwave} development kit,
the Tx/Rx sub-systems are fed by the same local oscillator and thus the \ac{cfo} is very low ($< 100$~Hz), which enables the extraction of the \ac{md} values required by SPARCS.  

\begin{table}[t!] 
\footnotesize
	\caption{Summary of the SPARCS implementation parameters. The suggested values based on experimental results are shown in bold. \label{tab:params}}
	\begin{center}
		\begin{tabular}{lcr}
			\toprule
			\multicolumn{3}{c}{{\bf System parameters }} \\
			\midrule
			Grid step & $T_c$ & $0.27$~ms\\
			Window length & $W$ & $64$\\
			Window shift & $\delta$ & $32$\\
			Sparsity parameter & $\Omega$ & $\{1,2,\boldsymbol{3},4,5,6,7\}$\\
			No. aggregated paths & $Q$ & $\{1,3,5,7,\boldsymbol{9},11,13,15\}$ \\ 
			Min. no. measurements & $M_s$ & $\{4, \boldsymbol{8}, 16, 24, 32, 64\}$\\
			\ac{iht} learning rate & $\eta$ & $1$\\
			\ac{iht} convergence threshold & $\xi$ & $10^{-4}$\\
			\ac{iht} maximum iteration number & $n_{\rm max}$ & $200$\\
			\bottomrule
		\end{tabular} 		
	\end{center}
\end{table}
\begin{figure*}[t!]
	\begin{center}   
		\centering
		\subcaptionbox{Ground truth \ac{stft}.\label{fig:gt}}[2.8cm]{\includegraphics[width=2.7cm]{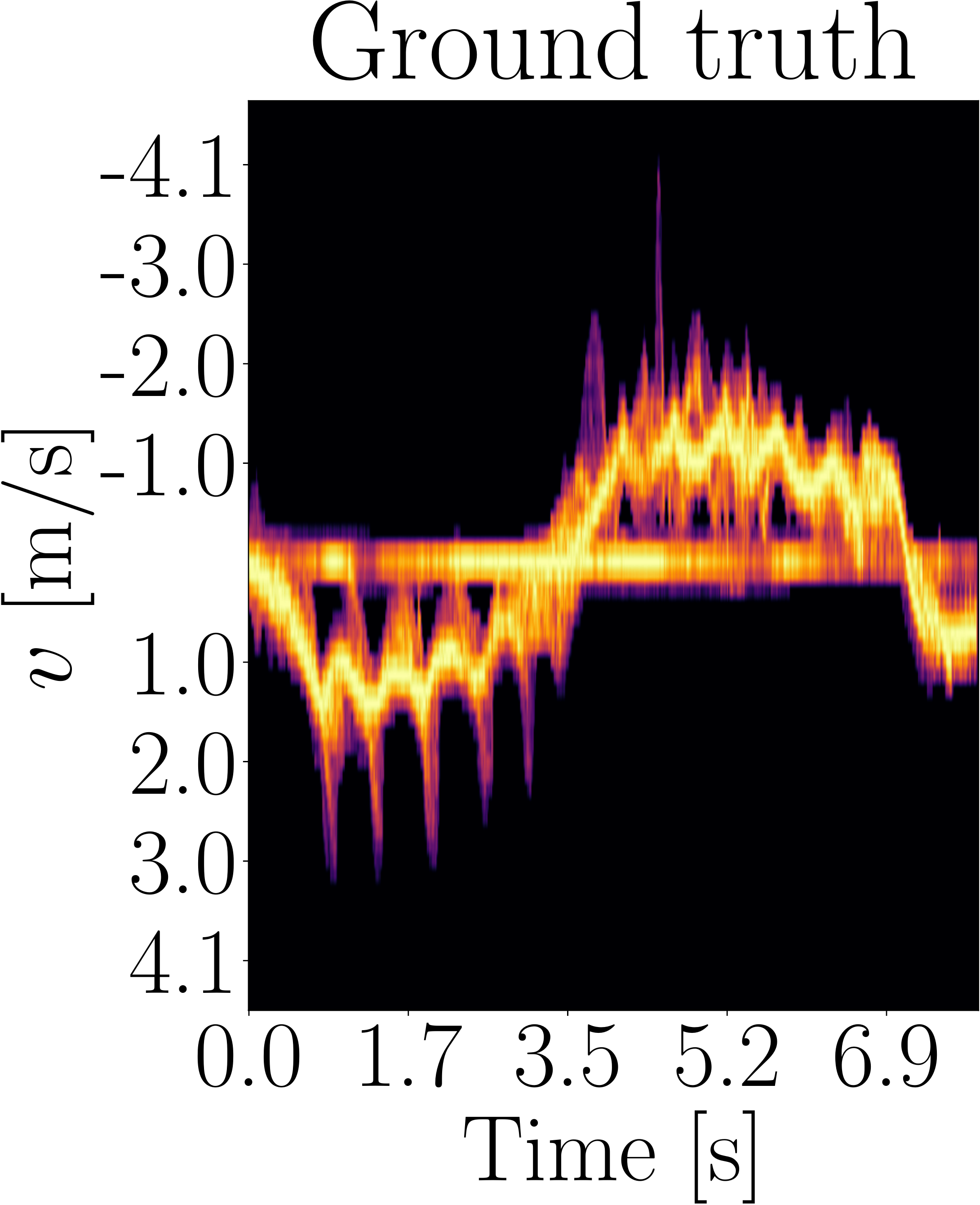}}
		\subcaptionbox{Full window (\ac{stft}).\label{fig:stft-64}}[2.8cm]{\includegraphics[width=2.7cm]{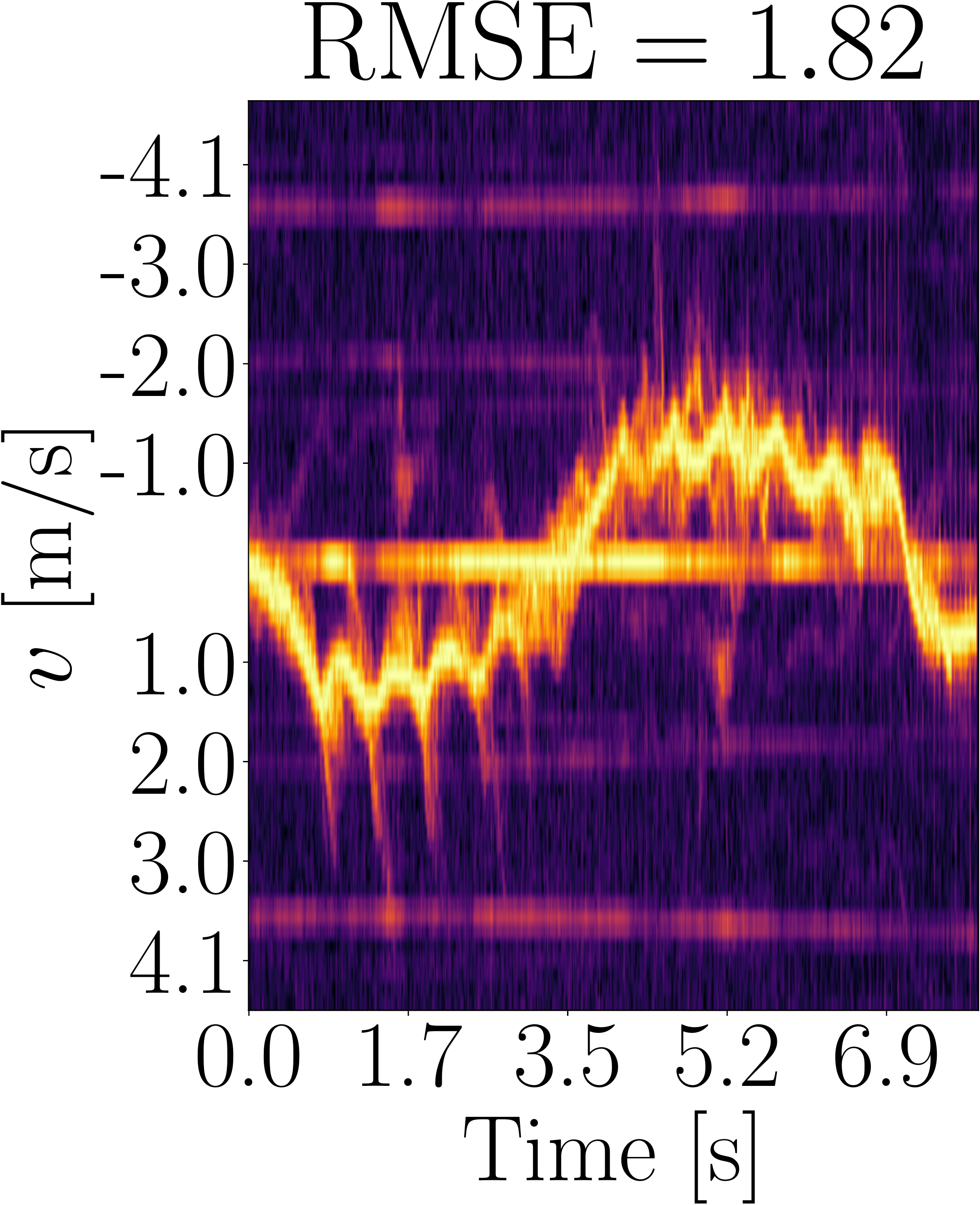}}
		\subcaptionbox{$1/4$ sparse (\ac{stft}).\label{fig:stft-sparse-64}}[2.8cm]{\includegraphics[width=2.7cm]{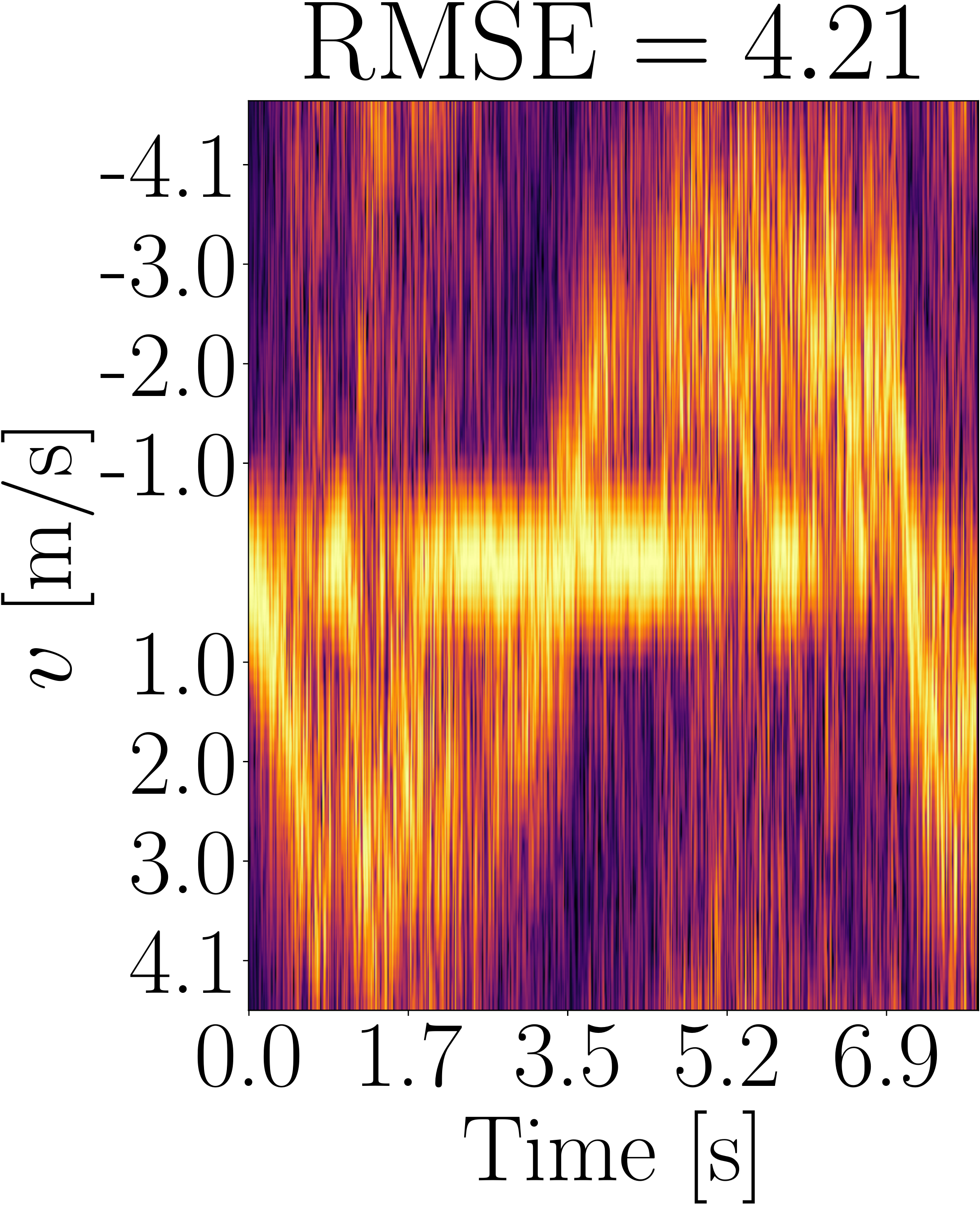}}
		\subcaptionbox{Full window (SPARCS).\label{fig:sparse-64}}[2.9cm]{\includegraphics[width=2.7cm]{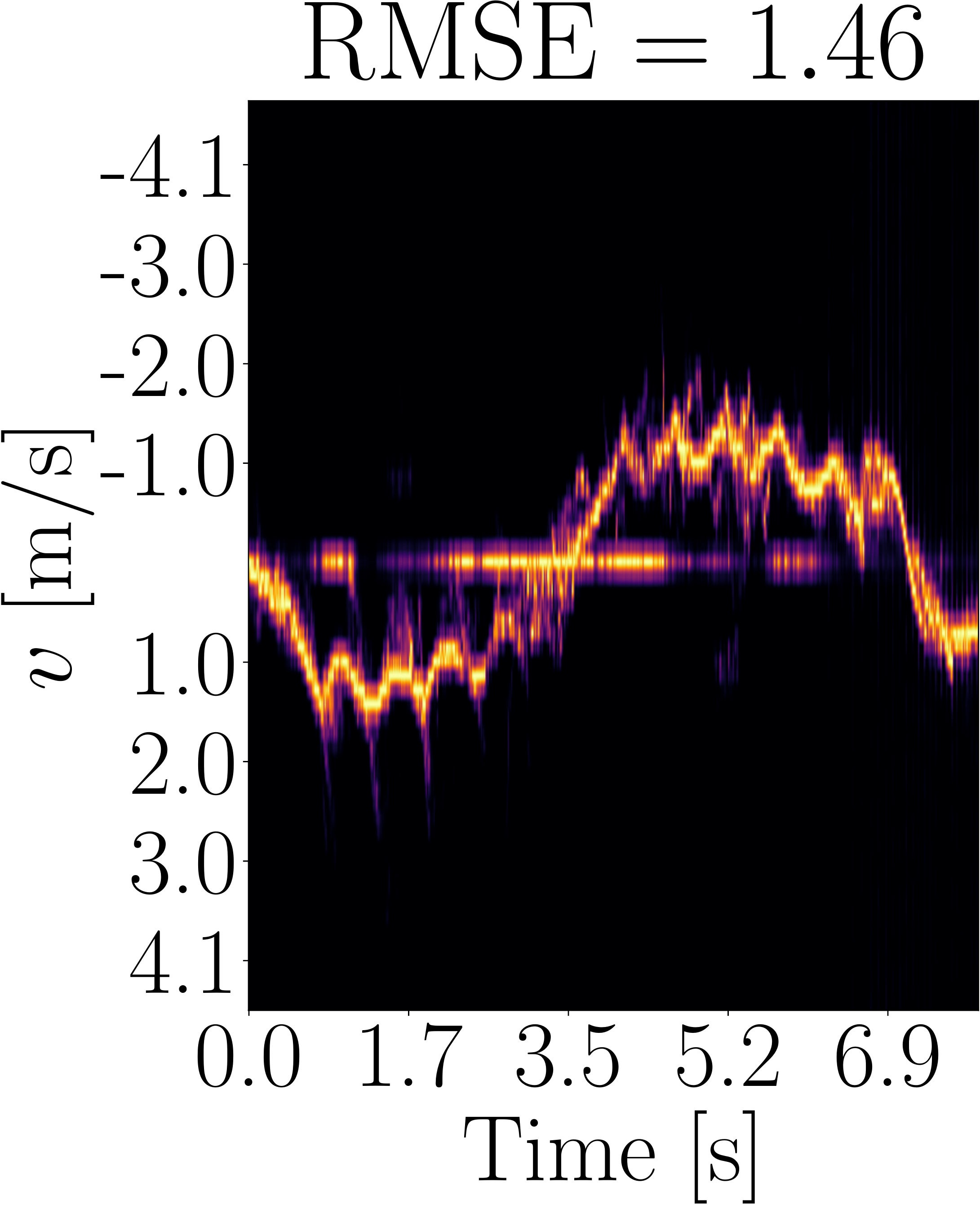}}
		\subcaptionbox{$1/4$ sparse ( SPARCS). \label{fig:sparse-16}}[2.8cm]{\includegraphics[width=2.7cm]{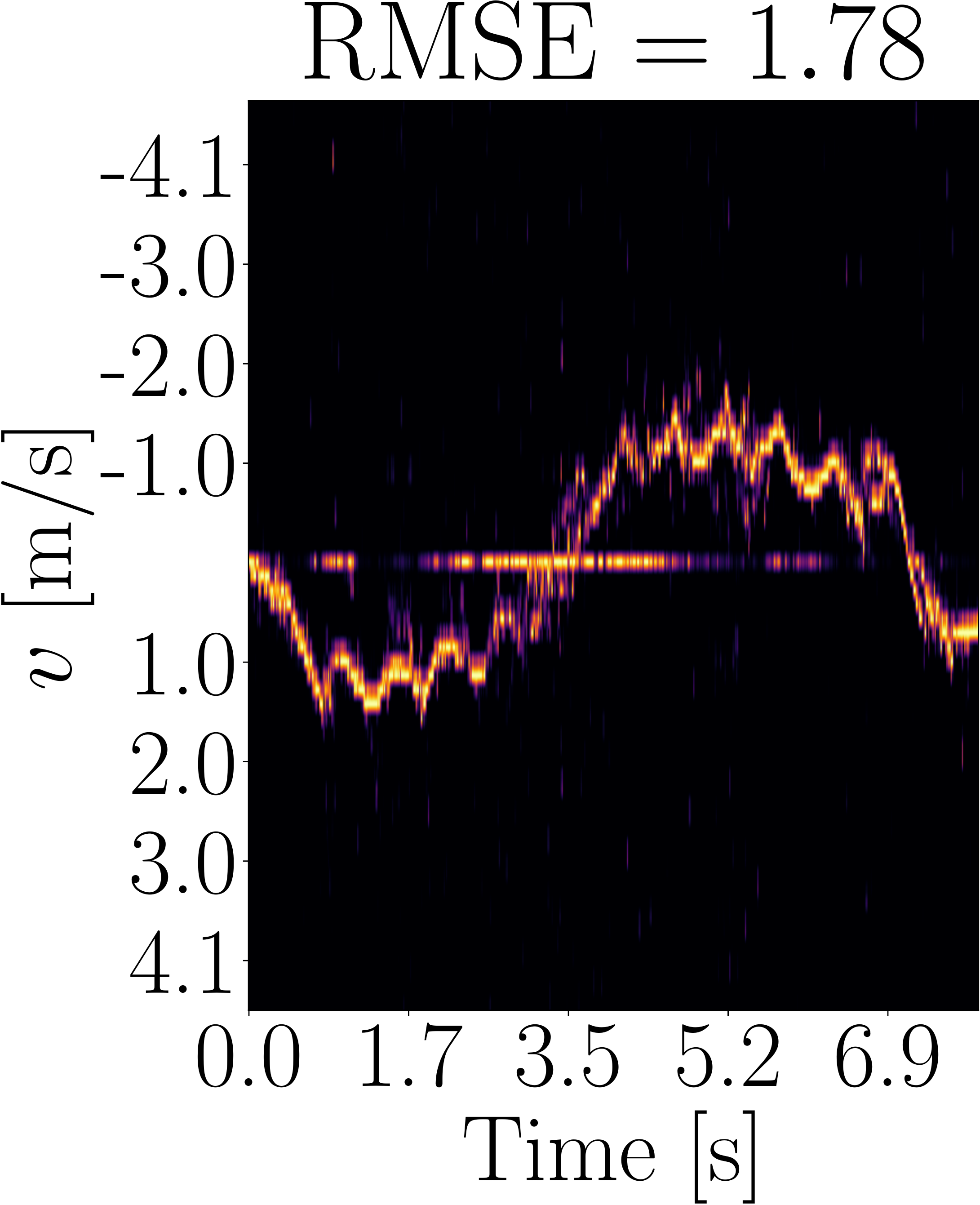}}
		\subcaptionbox{$1/16$ sparse (SPARCS). \label{fig:sparse-4}}[3.2cm]{\includegraphics[width=3.198cm]{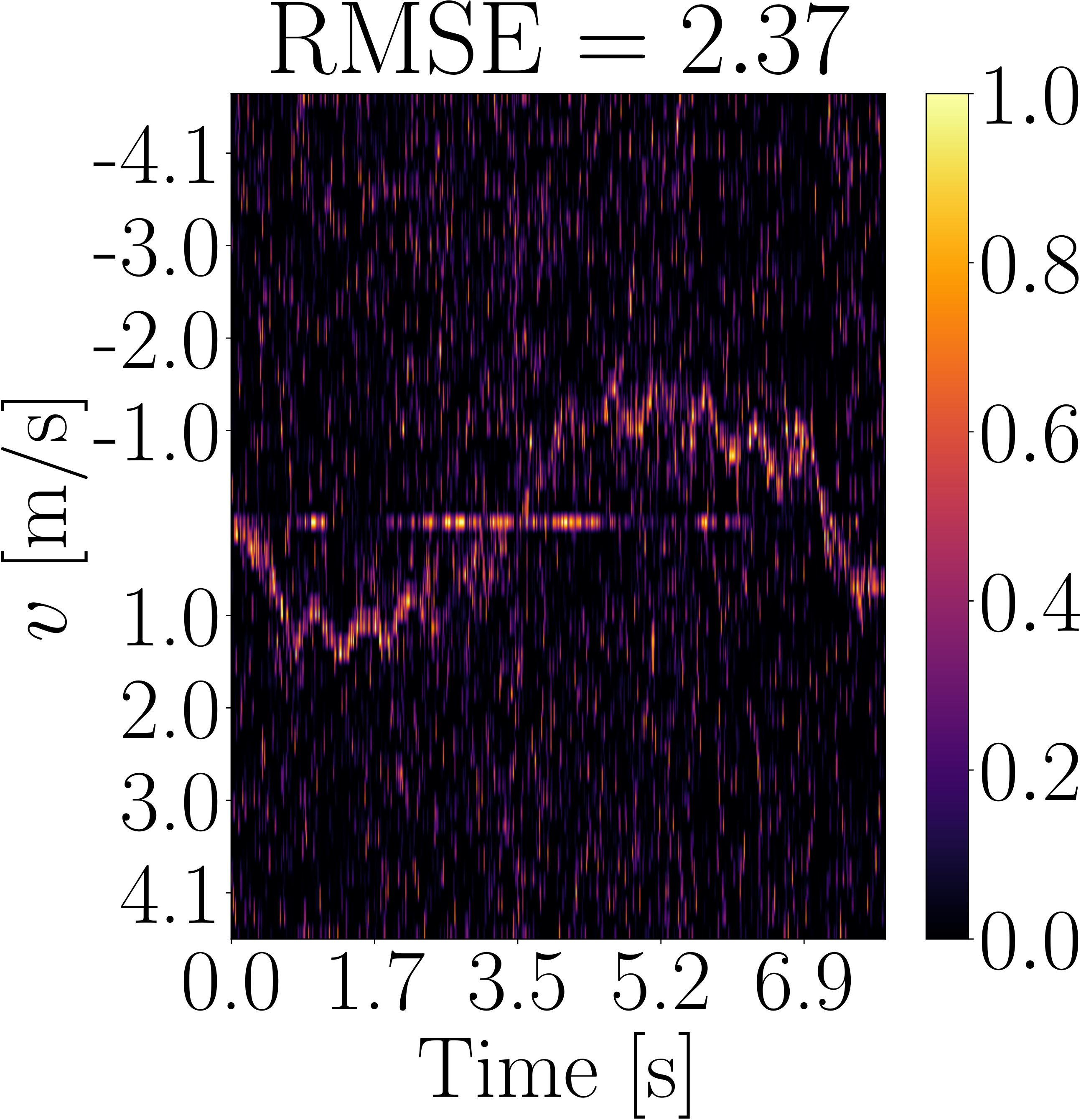}}
\vspace{-0.8em}
		\caption{\rev{Walking \ac{md} spectrograms and RMSE for different levels of sparsity, obtained by uniformly removing samples for each window.}}
		\label{fig:sparse-test}
	\end{center}
\end{figure*}

\noindent \textbf{IEEE~802.11ay \ac{cir} estimation details.}
\rev{In IEEE~802.11ay, \textit{in-packet} beam tracking \cite{Knightly_ICOMM2017} is introduced, where the \ac{cir} is estimated using different \acp{bp} \textit{within a single packet}. This is implemented by appending a given number of training (TRN) fields to the packet. A TRN field is composed of $6$ TRN \textit{units} formed by complementary Golay sequences of $128$~BPSK modulated samples, for a total of $768$~samples \cite{802.11ay}. 
In our implementation, we use $n_{\rm TRN}$ TRN fields as the SPARCS sensing unit, where each TRN field employs a different \ac{bp}, and $n_{\rm TRN}$ is the number of subjects being tracked by the system, as a single TRN field per subject suffices. Considering the typical number of people that are to be simultaneously tracked in human sensing systems, reasonable $n_{\rm TRN}$ values range from $1$ to $10$. The \ac{cir} estimates obtained from the TRN fields are then used as the input to SPARCS sparse recovery algorithm.}

\noindent \textbf{System parameters.} In \tab{tab:params} we summarize the system parameters used in the implementation.
We set $T_c = 0.27$~ms and $W=64$, which lead to \textit{(i)} a velocity resolution of $\Delta v= c/(2f_o W T_c)\approx 0.14$~m/s and \textit{(ii)} aliasing-free velocity measurements up to $v_{\rm max}= \pm c/(4f_o T_c)\approx \pm 4.48$~m/s. These values are not critical to the functioning of our system, and can be modified according to specific implementation requirements. However, for reliable \ac{md} extraction without aliasing, it is advisable to adjust $T_c$ to a value that allows capturing the range of velocities typically covered by human movement, e.g., approximately $\pm 2-3$~m/s for a walking person, and up to $\pm 5$~m/s for running or other fast movements \cite{vandersmissen2018indoor}. 
Note that suitable values of $T_c$ can also be obtained in $5$G-NR systems, where a base station can transmit downlink \ac{csirs} frames with a periodicity between $0.3125$~ms and $80$~ms. For a $5$G-NR carrier frequency of $28$~GHz, using $T_c=0.3125$~ms leads to $v_{\rm max}\approx \pm 8.57$~m/s, which is enough to capture fast human movement.

\rev{For people tracking, we 
use periodically transmitted in-packet beam training frames with $12$ TRN units and antenna beams covering a \ac{fov} range from $-45^\circ$ to $45^\circ$. Then, we utilize the distance and \ac{aoa} estimation procedure described in \secref{sec:track}, as proposed in \cite{pegoraro2021rapid}, to which we refer for further details}. 
We experimented with different values of $M_s, \Omega$ and $Q$, as reported in \tab{tab:params} and described in \secref{sec:har} and \secref{sec:params-sens}, while for the \ac{iht} algorithm we selected the parameters that led to the most accurate convergence results on our experiments, i.e., $\eta=1$, $\xi=10^{-4}$ and $n_{\max}=200$.

\section{Experimental results} \label{sec:results}
We now present the experimental results obtained with our SPARCS testbed implementation.
The experiments were performed in a laboratory of $6$ $\times$ $7$~meters with a complex multi-path environment due to additional reflections caused by furniture, computers, screens, and a wide whiteboard. 

\subsection{Results on synthetic traces}
\label{sec:qualitative}

As a first qualitative result we show the \ac{md} spectrograms obtained by SPARCS on randomly sampled \acp{cir} of a walking subject (see \fig{fig:sparse-test}). For this, we use synthetic traces, generated by measuring the \ac{cir} using a uniform sampling interval equal to $T_c$, and then setting to $0$ a variable number of uniformly distributed values \textit{per window} to simulate missing samples. This is a simplified case, as \textit{(i)} the available (not removed) packets lie on a regular grid with spacing $T_c$, therefore no approximation error is introduced by slotted resampling, and \textit{(ii)} samples are removed on a per window basis, so a minimum number of packets in each window is guaranteed. Still, this evaluation is useful to highlight the impact of increasing the sparsity level of the measurements for SPARCS compared to standard \ac{stft} \cite{pegoraro2021multiperson}.
In the results presented in this section, no packet injection is performed, as we aim to assess the impact of the number of measurements per window on the reconstructed \ac{md}.
In \fig{fig:stft-64} we show the baseline walking spectrogram obtained using the standard \ac{stft} using the full window of $64$ samples, as done in \cite{pegoraro2021rapid}. The spectrogram shows a typical walking \ac{md} modulation, with the contribution of the static clutter (the strong component at $0$ velocity), of the torso (the strong oscillating component around $\pm 1.5$~m/s and the limbs (the faint contributions around the torso component). Moreover, a certain amount of noise and interference is present, as shown by the non-zero background level and the horizontal lines at around $\pm 2$~m/s and $\pm 3.7$~m/s. In \fig{fig:stft-sparse-64}, the same method is applied to windows with only 16 out of 64 the samples retained, while the rest is set to $0$. The impact is very strong as it completely corrupts the useful structure in the \ac{md} signature.
From \fig{fig:sparse-64} to \fig{fig:sparse-4} we show the results obtained by SPARCS, on the same sequence, with $64, 16$ and $4$ samples out of $64$, respectively. 
\rev{At the top of each figure, we report the Root Mean-Squared Error (RMSE) of the \ac{md} with respect to a ground truth spectrogram, shown in \fig{fig:gt}. This ground truth was obtained from the \ac{stft} output with full measurement windows (\fig{fig:stft-64}), by manually isolating the useful \ac{md} spectrum containing the gait information and setting to $0$ any background noise and interference lines.}
We observe two interesting aspects. On the one hand, the SPARCS algorithm can successfully recover the \ac{md} spectrum even when a large fraction of the samples is missing, and the quality of the result decreases \textit{gracefully} with the sparsity of the available measurements. \rev{Unlike standard \ac{stft}, SPARCS almost completely eliminates the noise and interference in the estimated $\mu$D spectrogram. Such improvement is made possible by the sparsity constraint in \eq{eq:local-opt}, which allows for a lower RMSE than \ac{stft} operating on full measurement windows}. This is the main reason why SPARCS not only reduces the overhead needed for human sensing, but also improves its accuracy.

\subsection{Realistic traces: the \texttt{pdx/vwave} dataset}\label{sec:pdx-vwave}

\begin{figure}
     \centering
     \includegraphics[width=8cm]{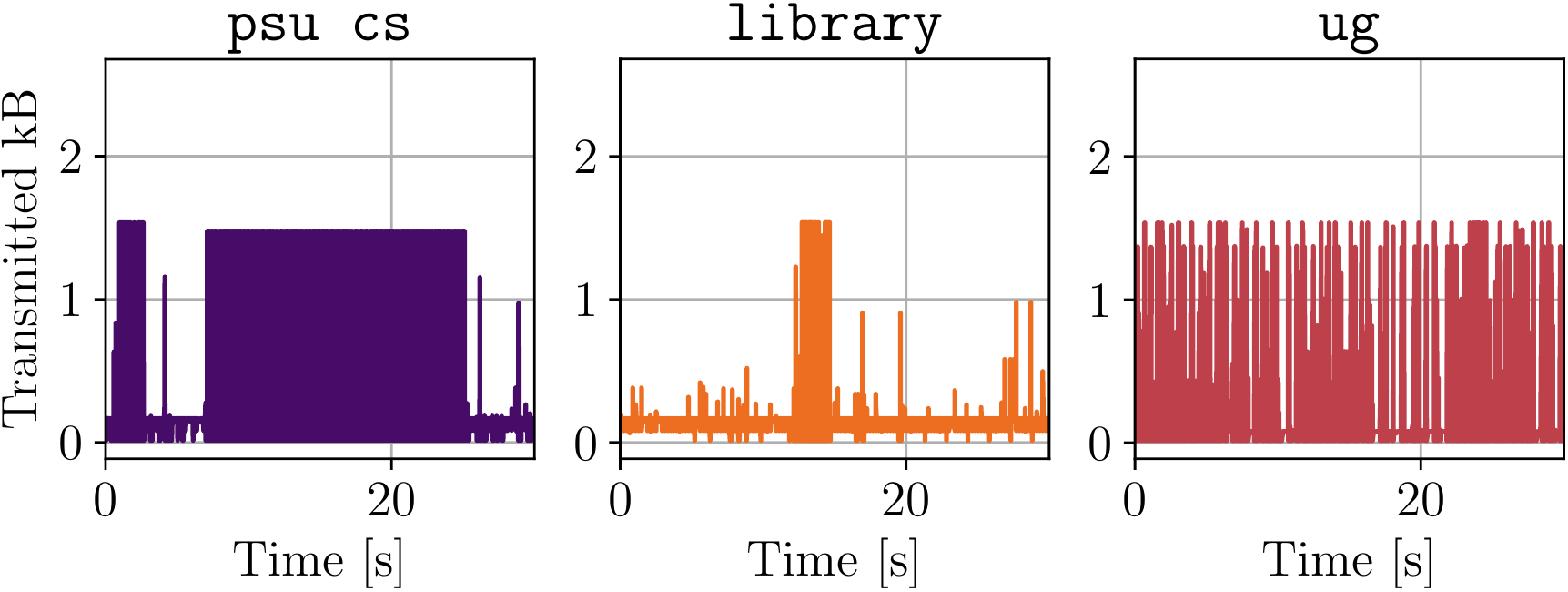}
\vspace{-0.5em}
     \caption{Example traffic patterns from the \texttt{pdx/vwave} dataset.}
     \label{fig:crawdad}
\end{figure}
\begin{table}[t!] 
\footnotesize
	\caption{Details of the $3$ sequences of the \texttt{pdx/vwave} dataset.} \label{tab:pdx-vwave}
    \vspace{-0.25cm}
	\begin{center}
		\begin{tabular}{lccc}
			\toprule	
			\textbf{Trace} & \textbf{Environment} & \textbf{No. frames} & \textbf{Duration}\\
			\midrule
			\texttt{psu cs} & University CS dept.&$260326$&$1:00$~h\\
			\texttt{library} & Public library&$1300671$&$4:00$~h\\
			\texttt{ug} &Coffee shop &$895721$&$2:34$~h\\
			\bottomrule
		\end{tabular} 		
	\end{center}
\end{table}
Next, we evaluate the performance of SPARCS on realistic WiFi \ac{ap} traces. This poses an experimental challenge, because commercial devices implementing the IEEE~802.11ay standard are not yet available, and no public datasets containing real traffic traces for the \ac{phy} layer of \ac{mmwave} WiFi (IEEE~802.11ay/ad) exist, to the best of our knowledge.
For this reason, we used the \texttt{pdx/vwave} dataset, containing real traffic traces captured in different real environments from WiFi \acp{ap} employing a legacy (sub-$6$~GHz) WiFi protocol \cite{pdx-vwave-20090704}.
Specifically, we use $3$, over $1$ hour long, traces from this dataset, called \texttt{psu cs}, \texttt{library} and \texttt{ug}, respectively. We select traces collected in different environments to represent different kinds of traffic patterns (see \tab{tab:pdx-vwave}).

The \texttt{pdx/vwave} dataset includes information about the transmission instants and packet sizes of all packets outgoing from the considered \ac{ap}. Exploiting this information, we perform our measurements transmitting packets according to these time patterns (see \secref{sec:har}), using the BRAM in the FPGA to store the desired transmission instants (see \secref{sec:implementation}).
On top of the existing \texttt{pdx/vwave} communication patterns we use the injection algorithm (\alg{alg:inj-algo}) to send additional sensing units when needed.

Even though the \texttt{pdx/vwave} dataset is based on a legacy sub-6 GHz WiFi protocol,
we argue that it is still reasonable to use it to obtain realistic packet transmission patterns.
While in the \texttt{pdx/vwave} dataset the maximum physical layer PDU size is $\mathrm{PPDU}_{\rm pdx} = 1.5$~kB (without packet aggregation), in IEEE~802.11ay three main transmission modes are defined, namely \ac{ht}, \ac{dmg} and \ac{vht}, with maximum physical layer PDU sizes, $\mathrm{PPDU}_{\rm ay}$, of $65$~kB, $262$~kB and 
$4692$~kB, respectively \cite{802.11ad, 802.11ay}. With the increase in the packet sizes, the data rates of \ac{mmwave} systems have increased accordingly, and in IEEE~802.11ay they will range from $0.3$~Gbps to several Gbps. As a numerical example, the traffic patterns in \texttt{pdx/vwave} with a typical bitrate of $4$~Mbps would correspond to a bitrate of $0.7$~Gbps in \ac{dmg} IEEE~802.11ay when using an aggregated packet size of $262$~kB instead of $1.5$~kB. Note that traces with a larger number of packets and smaller PDU sizes (as will likely be the case in real deployments) will simply increase the sensing accuracy and further reduce the overhead.

\subsection{Human activity recognition results}\label{sec:har}
To evaluate the quality of the \ac{md} spectrograms extracted by SPARCS, we use them as the input to a \ac{har} method. Specifically, we follow a standard approach, training a deep neural network on a dataset of $\Lambda \times W$ dimensional \ac{md} spectrograms, with $\Lambda = 200$ (equivalent to $\approx 1.76$~s), in order to classify the movement performed by the person during that time. In order to provide a comparison with other IEEE~802.11ay \ac{har} methods based on regular \ac{cir} sampling, such as RAPID \cite{pegoraro2021rapid}, we consider the $4$ following activities: walking, running, sitting and waving hands.
For \ac{har}, we use a standard \ac{cnn} architecture, composed of $4$ \textit{inception modules} \cite{szegedy2015going} performing $1\times 1$, $3\times 3$ and $5\times 5$ convolutions. The number of filters used is $8, 16, 32$ and $64$ for the $4$ modules, respectively. The convolutional blocks are followed by a fully-connected layer with $64$ neurons, to which we apply Dropout \cite{srivastava2014dropout}, and a final Softmax layer with $4$ outputs \cite{goodfellow2016deep}. We use the \textit{exponential-linear unit} activation function after each layer \cite{clevert2015fast}. 

\subsubsection{Training data.} We collected a training dataset involving $6$ different subjects performing the $4$ activities, for a total duration of about $12$ minutes each. This leads to over $400$ partially overlapping, $1.76$~s long, \ac{md} sequences per activity, which we then augmented as described shortly. Note that the training data only includes \textit{uniformly sampled} \ac{cir} traces with sampling period $T_c = 0.27$~ms.
The \ac{cnn} training is done for $80$ epochs, using a learning rate of $10^{-4}$, the Adam optimizer and the cross-entropy loss function \cite{goodfellow2016deep}. 
In order to enhance the robustness of the \ac{cnn}, we apply an ad-hoc data augmentation strategy: we randomly remove some of the \ac{cir} samples in each window of the training dataset, and then apply SPARCS' \ac{iht} algorithm to reconstruct the spectrograms (see \secref{sec:qualitative}). We repeat the process using a sparsity level of $1/8, 1/4$ and $1/2$, enlarging the training dataset to $4$ times its original size, for a total of approximately $1600$ \ac{md} spectrograms per activity. \rev{A randomly selected subset of the training data (around $10$~\%) was used as a validation set to tune the \ac{cnn} hyperparameters.}

\subsubsection{Test data.} We test the \ac{cnn} on the \ac{md} spectrograms obtained from \ac{cir} samples collected using the \texttt{pdx/vwave} packet traces described in \secref{sec:pdx-vwave}. We collect four, randomly selected, $20$~s long traces (one per activity) for each of the $3$ sequences types (\texttt{psu cs}, \texttt{library} and \texttt{ug}). We repeat the experiments for different values of the minimum number of sensing units per window, $M_s = 4, 8, 16, 32, 64$, for a total of $60$ test sequences. The test data involves a single subject, which was not included in the training set.

\begin{figure}
     \centering
     \includegraphics[width=8.4cm]{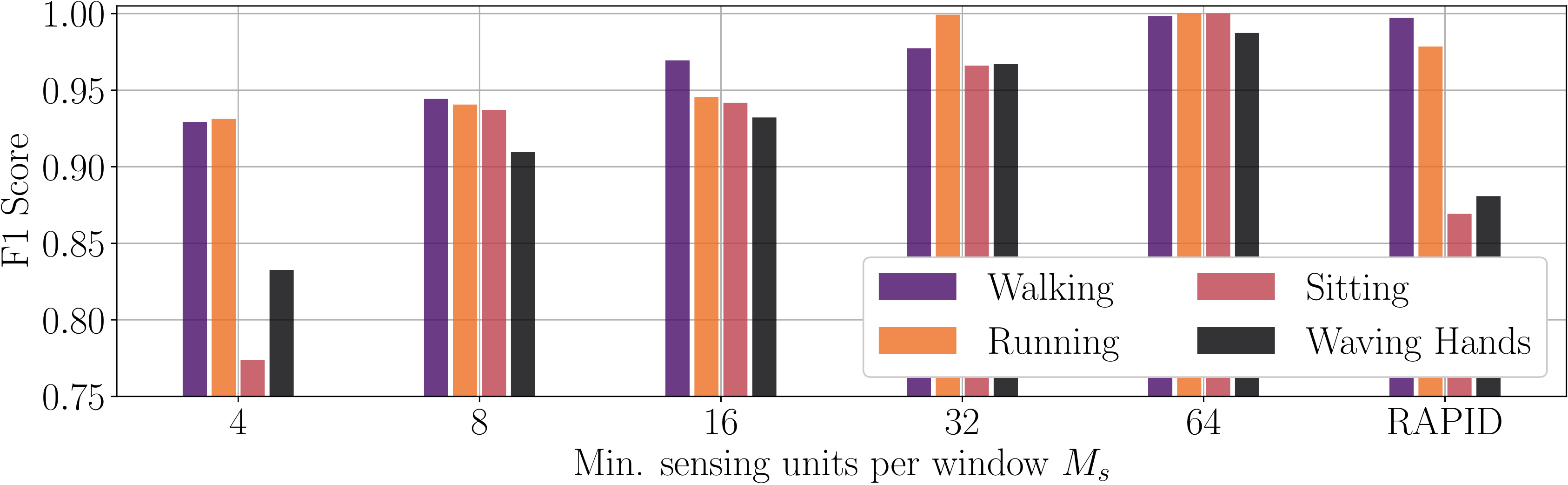}
     \caption{Per-class F$1$ scores obtained by SPARCS (for different values of $M_s$) and RAPID on our test dataset \cite{pegoraro2021rapid}.}
     \label{fig:f1scores}
\end{figure}
\subsubsection{\ac{har} F$1$ score.} We evaluate the performance of the \ac{cnn} with the \textit{per-class} F$1$ score metric \cite{chinchor1992muc}, which effectively summarizes the precision and recall and preserves the class-specific results. 
\fig{fig:f1scores} shows the total average per-class F$1$ score over the $60$ sequences, for different values of the minimum number of sensing units per window, $M_s$. As a baseline for comparison, we also report the F$1$ score obtained by the RAPID algorithm from \cite{pegoraro2021rapid}, which extracts the \ac{md} signatures by regularly sampling the \ac{cir}.  Our results show that SPARCS can reach over $0.9$ F$1$ scores on all activities with $M_s=8$ already, which corresponds to only $1/8$ of the full measurements window. Notably, with $M_s=4$, the low number of measurements per window affects significantly only the 'Sitting' and 'Waving hands' activities, which involve fine-grained movements and are therefore more difficult to classify.
Finally, we compare the results from SPARCS and RAPID \cite{pegoraro2021rapid}. For a fair comparison, we implemented RAPID's \ac{stft} to extract the \ac{md} and trained the \ac{cnn} on the resulting spectrograms without enlarging the dataset using different levels of sparsity described in the previous section. Instead, we directly use the training procedure of \cite{pegoraro2021rapid}, since we found that the sparsity-based data augmentation slightly reduced RAPID's performance. It can be seen that SPARCS' sparse recovery problem formulation (\secref{sec:sparse-rec}) and enforcing a sparsity constraint on the individual paths is beneficial to \ac{har} performance. The gap is particularly significant for 'Sitting' and 'Waving hands' as they involve lower energy traces in the spectrograms; these are more easily corrupted by noise and interference, that SPARCS is mostly able to reject (see, again, the comparison between \fig{fig:stft-64} and \fig{fig:sparse-64}). 

\subsection{Overhead analysis}
\label{sec:overhead-res}

Increasing $M_s$ to improve the \ac{har} performance also increases the overhead of SPARCS.
A first general measure of this can be obtained comparing the maximum size of a PPDU in IEEE~802.11ay to the size of a sensing unit. Recalling the three different modes introduced in \secref{sec:pdx-vwave} and the size of an IEEE~802.11ay TRN field ($768$~bits), we obtain that a sensing unit, with $n_{\rm TRN} = 1$, is $0.1\%$, $0.03\%$ and $0.002\%$ of a PPDU in \ac{ht}, \ac{dmg} and \ac{vht}, respectively. 
Moreover, the channel occupation time for a sensing unit with $n_{\rm TRN} = 1$ is $436$~ns \cite{802.11ay}, which is a negligible fraction ($0.16\%$) of a slot of duration $T_c$.

\begin{figure}[t!]
	\begin{center}   
		\centering
		\includegraphics[width=4.3cm]{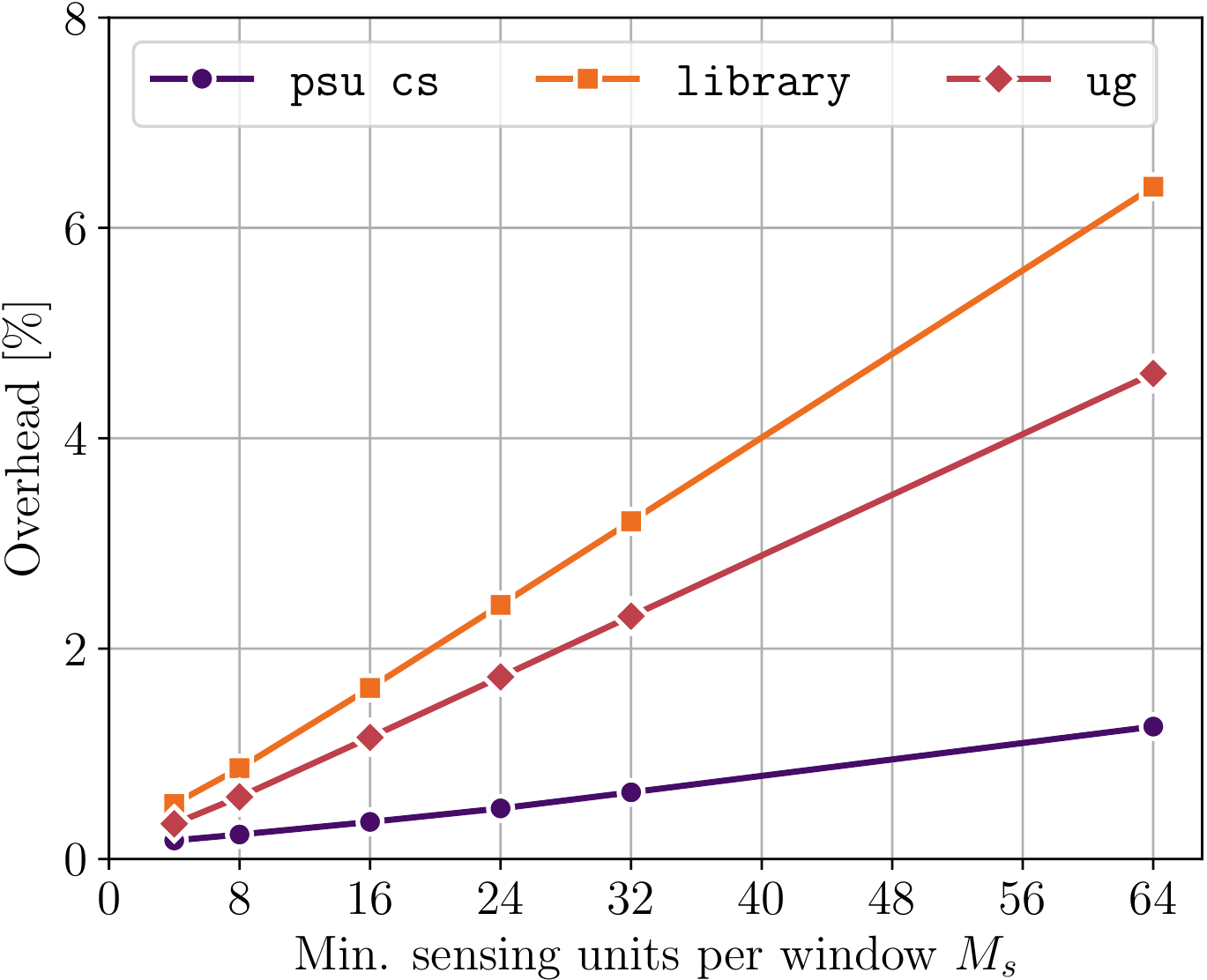}
		\includegraphics[width=4cm]{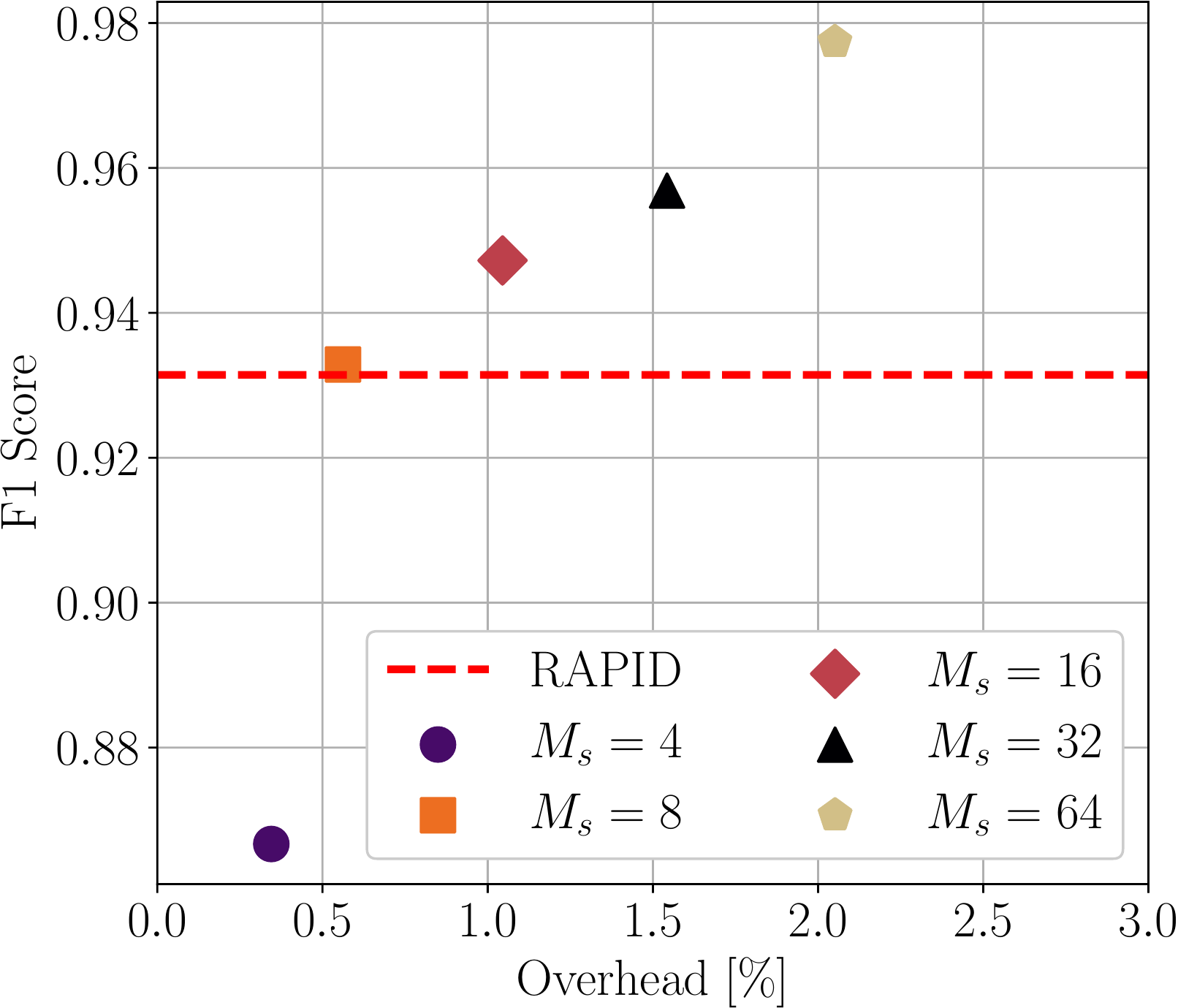}
		\caption{Overhead of SPARCS for different values of $M_s$ in the three traces of the \texttt{pdx/vwave} dataset (left). Overhead vs.\ average \ac{har} F$1$ score for different values of $M_s$ (right). 
		}
		\label{fig:overhead}
	\end{center}
\end{figure}

Next, to evaluate the overhead of SPARCS on a realistic communication scenario, we use the traces of the \texttt{pdx/vwave} dataset. In this way, we can also assess the impact of injecting sensing units, as they are not useful to the communication process.
Denote by $c_i$ the number of bits in the $i$-th communication packet transmitted in the trace.
As the number of bits transmitted in each trace refers to a legacy, lower bitrate, WiFi protocol, we rescaled the packet sizes according to the maximum PHY layer packet size in IEEE~802.11ay. 
We rescaled the size of packet $i$ in each trace as $\tilde{c}_i =  (\mathrm{PPDU}_{\rm ay} / \mathrm{PPDU}_{\rm pdx}) \times c_i$, with $\mathrm{PPDU}_{\rm ay} = 262$~kB.
We call $n_{c}$ the number of transmitted communication packets in a trace, by $\mathrm{TRN}_{\rm len}$ the length, in bits, of a piggybacked or injected TRN field, $n_{\rm inj}$ the number of injected sensing units and $n_{\rm TRN}$ the number of TRN fields used in every sensing operation (we consider it fixed, whereas in reality it is determined by the number of subjects in the environment). We define the overhead as a function of $n_{\rm inj}$ as 
\begin{equation}
\mathrm{OH}(n_{\rm inj}) = \frac{n_{\rm TRN}\left(n_{C} + n_{\rm inj}\right)\mathrm{TRN}_{\rm len}}{\sum_{i=1}^{n_{c}}\tilde{c}_i}.
\end{equation}
In \fig{fig:overhead}, left, we show the overhead obtained on each of the three \texttt{pdx/vwave} traces, using $n_{\rm TRN} = 1$. The overhead for different values of $n_{\rm TRN}$ can be obtained by using it as a multiplicative factor on the values in \fig{fig:overhead}. We see that the overhead scales almost linearly as $M_s$ is increased from $4$ to $64$. 
For values of $M_s < 32$, the entailed overhead is less than $4$\%, falling below $1$\% for $M_s=8$
As a reference, we report the overhead for $M_s=64$, which is the value obtained by injecting sensing units continuously into the channel, piggybacking them eventually on communication packets if possible. Note that existing approaches requiring uniform \ac{cir} sampling, like RAPID \cite{pegoraro2021rapid}, would require an even higher overhead, as not only do they need $64$ samples per window, but these samples have to be regularly spaced as no resampling procedure is carried out. This means they would have to take precedence over potential data packets so that they are sent exactly at the right sampling time.

From \fig{fig:overhead}, right, one can see that SPARCS can achieve an F$1$ score of over $0.9$ for every activity for a minimum of $M_s=8$ sensing units per window, resulting in a sensing overhead of less than $1$\%. With this configuration, SPARCS achieves a better F$1$ score than existing approaches, while reducing overhead by a factor of $7$ and being compatible with random access MAC protocols.

\subsection{Sensitivity to the choice of the parameters}
\label{sec:params-sens}
In \fig{fig:params-vary}, we show the effect of varying parameters $Q$, representing the number of paths aggregated around the person's position (see \eq{eq:sparse-paths-sum}), and $\Omega$, which is the maximum number of resolvable Doppler components, equal to the sparsity parameter in the \ac{iht} algorithm.
We computed the \ac{har} per-class F$1$ score using a random subset of the $60$ test sequences. The values adopted in our experiments are reported in \tab{tab:params}.

\subsubsection{Impact of changing $Q$} Our results show that SPARCS is robust to almost any value of $Q$ when considering walking and running, whereas sitting and waving hands are negatively affected by reducing $Q$ below $7$. This is due to the fact that while walking and running are, in most cases, distinguishable even from the sole contribution of the torso, this is not true for sitting and waving hands that require including the reflection paths coming from the limbs. Computational complexity considerations are also in order for high values of $Q$, as it leads to solving $Q$ times \eq{eq:local-opt} at each \ac{md} extraction process. As the problems are independent, they can be solved in parallel, and thus a reasonable approach is to tune $Q$ according to a trade-off between \ac{md} reconstruction accuracy and hardware resource availability for parallelization. 
In the following, we use $Q=9$. Considering that we use $B=1.76$~GHz transmission bandwidth ($1$ IEEE~802.11ay channel), the range resolution of SPARCS is $c/2B = 8.5$~cm. This means that summing the contribution of $\lfloor Q/2\rfloor$ distance bins before and after the one corresponding to the torso, we include in the spectrum a region of $\pm 34$~cm around the person's position, which is a reasonable value considering typical body sizes and that the subjects are moving. 

\subsubsection{Impact of changing $\Omega$} 
\begin{figure}
     \centering
     \includegraphics[width=4.6cm]{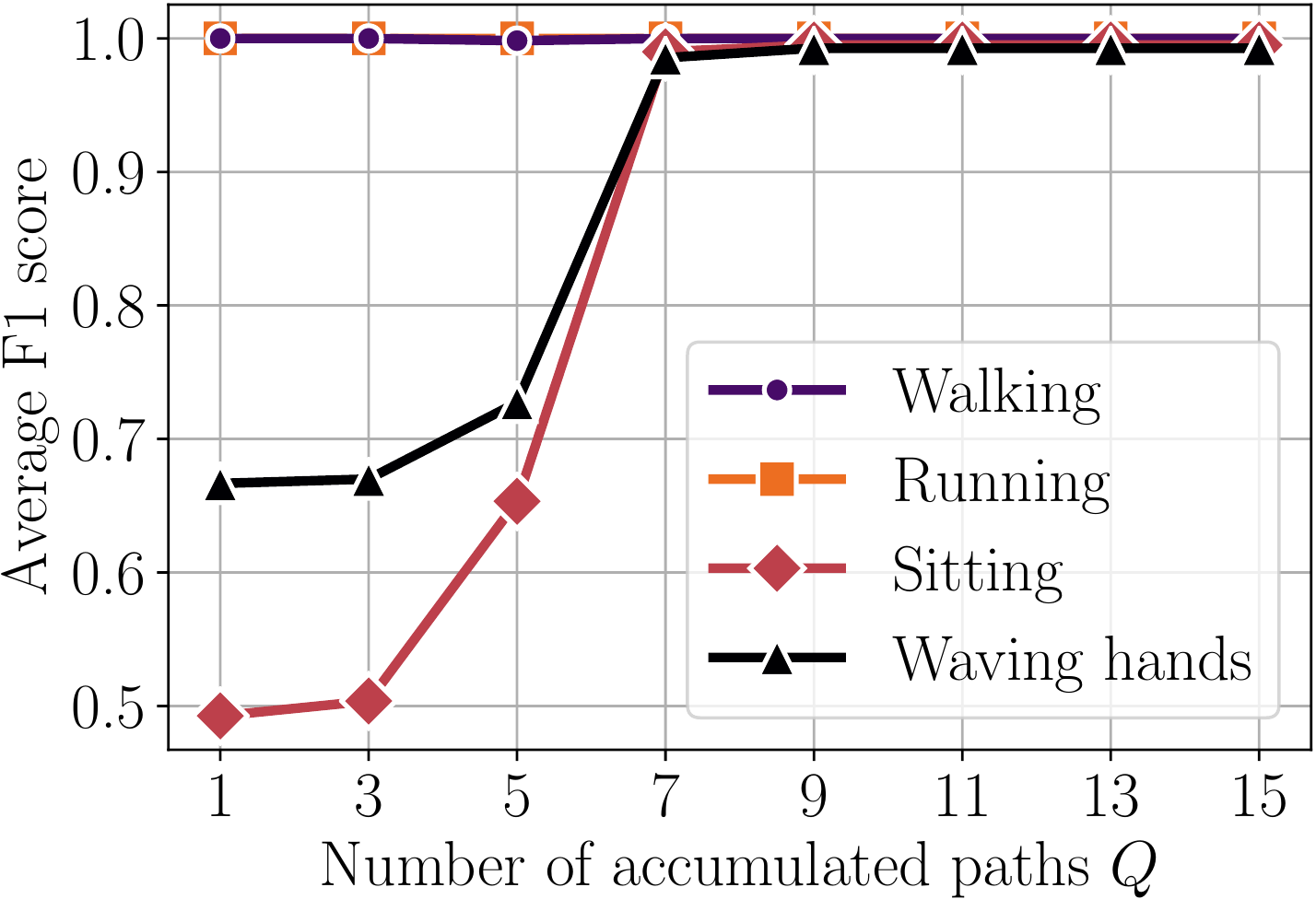}
	 \includegraphics[width=3.8cm]{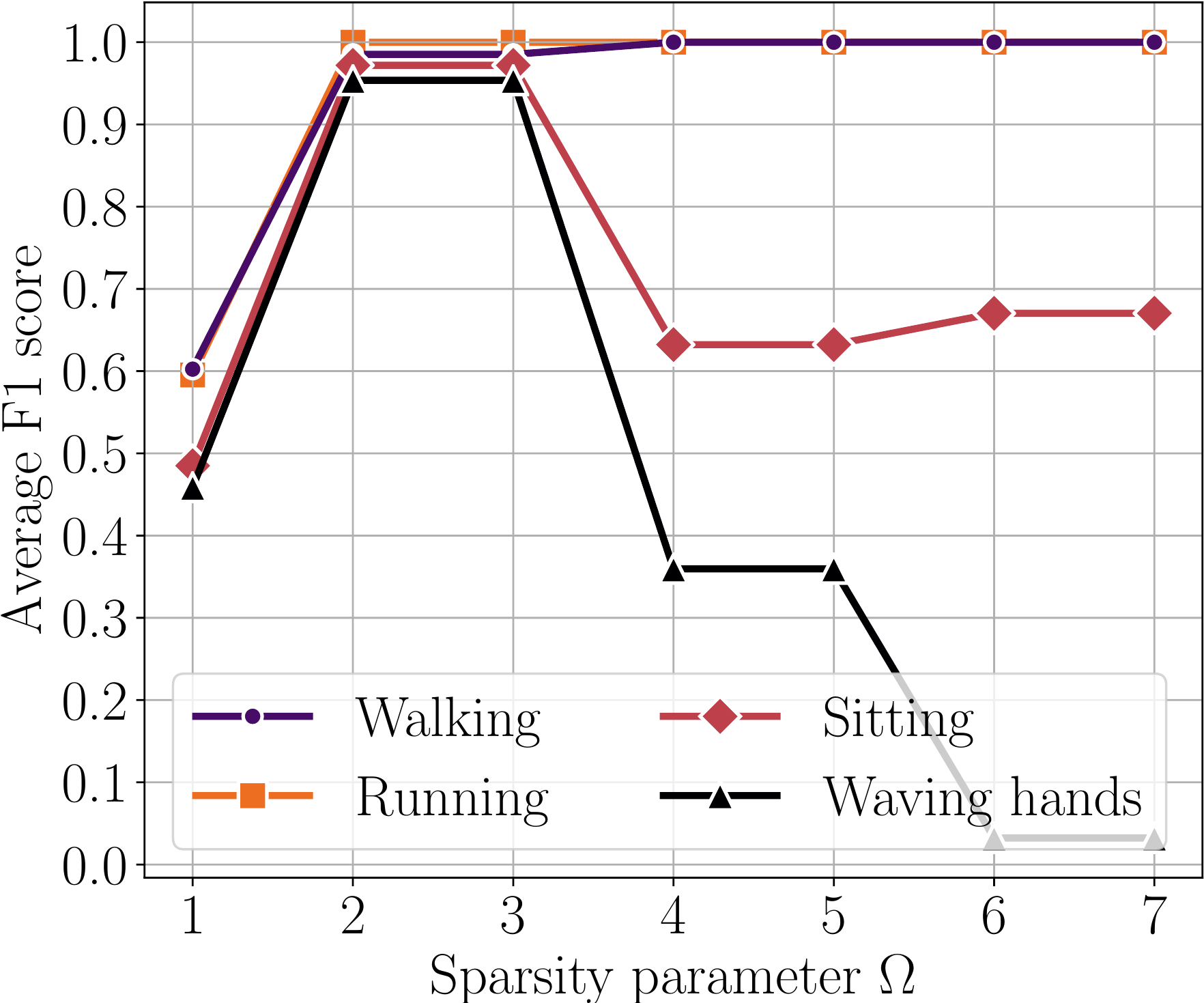}
     \caption{Per-class F$1$ scores aggregating a different number of paths $Q$ (left) and changing $\Omega$, the \ac{iht} sparsity parameter (right).}
     \label{fig:params-vary}
\end{figure}

Fixing $Q=9$, in \fig{fig:params-vary} (right), we show that the best values for $\Omega$ are $2$ and $3$ for all the activities. This is because using $\Omega=1$ often leads to only reconstructing the $0$ Doppler component in the spectrogram, losing the information on the person's movement. On the other hand, choosing $\Omega$ too high makes the \ac{iht} reconstruction imprecise, as with a low number of measurements per window enforcing more sparsity is beneficial to restrict the number of possible solutions to \eq{eq:local-opt}.

\section{Related work}\label{sec:rel}
\noindent\textbf{Dedicated \ac{mmwave} radars.} 
The high sensitivity of \acp{mmwave} to micro-Doppler shifts, together with \ac{dl} methods for spectrogram analysis and classification, have been widely exploited to enable applications such as activity recognition \cite{singh2019radhar, lai2021radar}, person identification \cite{meng2020gait, pegoraro2021multiperson} and bio-mechanical gait analysis \cite{seifert2019toward}.
The typical approach in these works is to transmit sequences of large bandwidth signals (of $2$ to $4$~GHz), with a rate dictated by the desired sensing resolution \cite{patole2017automotive}. Thus, \ac{mmwave} radar sensors have two main drawbacks:

\noindent\textit{(i)} they are specifically tailored to sensing and cannot perform communication. Moreover, their deployment cost is relatively high as one single sensor can reliably cover a range of at most $8-10$~m due to the radial distortion and occlusion problems  \cite{feng2020evaluating}. For this reason, ad-hoc radar sensor networks would need to be deployed in practical scenarios. Our method, in contrast, fully exploits existing \ac{mmwave} communication systems with no modifications to the \ac{cir} estimation process or packet structures.

\noindent\textit{(ii)} The fixed chirp transmission interval, which is related to the \ac{md} resolution, requires regular transmissions with continuous channel occupation. 
Some works have explored the possibility of randomly subsampling the chirp transmission intervals using compressive sensing \cite{eldar2012compressed} to either save computational resources \cite{stankovic2014compressive} or reduce the effect of unwanted interference \cite{sejdic2018compressive}.
However, these works are based on a radar framework, where the transmission instants can be freely chosen and optimized. SPARCS, instead, reuses the given underlying communication traffic as much as possible and only injects small additional sensing units when necessary.

\noindent \textbf{$\mathbf{60}$~GHz WiFi sensing.} Research interest towards sensing with WiFi devices working in the \ac{mmwave} band has mostly focused on the $60$~GHz IEEE~802.11ad/ay standards \cite{wu2020mmtrack, zhang2020mmeye, pegoraro2021rapid}. These works target various applications, such as person tracking and gesture recognition, exploiting \ac{cir} estimation to detect humans in the environment. However, they require dedicated and regular sensing signal transmissions in order to function properly, entailing a significant overhead and channel utilization for sensing. 

Our work significantly improves over the above mentioned studies in that it enables the reuse of randomly distributed communication packets via sparse recovery, whenever possible. This is of key importance to \emph{integrate} sensing capabilities in communication devices while maintaining low overhead and complexity.

\noindent \textbf{Integrated sensing and communication.}
A number of technical works address \ac{isac} systems in next generation $5$G/$6$G cellular networks \cite{liu2021integrated, li2021rethinking} and \acp{wlan} \cite{restuccia2021ieee, pegoraro2021rapid}. Many of those target the joint communication and sensing waveform design \cite{liu2018toward} and are mostly oriented to automotive applications to measure distance and velocity of nearby vehicles. 
In contrast, few works focus on human sensing \cite{li2021rethinking}, which is the aim of the present work. 
All of the above approaches alternate communication and sensing phases according to a time-division scheme, causing significant overhead and channel occupation. SPARCS instead, provides a full \ac{isac} scheme, as it passively exploits communication traffic while dynamically injecting sensing units to cover silent periods. As a result, our method  significantly reduces sensing overhead while at the same time improving the sensing accuracy. 

\section{Concluding remarks} \label{sec:conclusion}

In this paper, we have designed and implemented SPARCS, the first \ac{mmwave}  \ac{isac} system that can sense human \ac{md} signatures from irregular and sparse \ac{cir} estimates. These are obtained in a standard compliant way by both reusing optional \ac{cir} estimation fields appended to communication packets and sporadically injecting sensing packets whenever communication traffic is absent.  
Differently from the existing \ac{isac} methods, SPARCS is based on a sparse recovery approach to the \ac{md} reconstruction, which is theoretically grounded in the instrinsic sparse multi-path environment of the \ac{mmwave} channel. This enables an accurate \ac{md} extraction from a significantly lower number of randomly distributed \ac{cir} samples, thus drastically reducing the sensing overhead. After a \ac{cir} resampling step along the time domain, SPARCS performs an iterative sparse reconstruction in the frequency domain, decoupling different propagation paths at first, to leverage their sparsity property, and then combining them to obtain the final \ac{md} spectrum.

While SPARCS is compatible with different \ac{mmwave} systems (e.g., $3$GPP $5$G-NR, and IEEE $60$~GHz \acp{wlan}), for our implementation we used an IEEE~802.11ay \ac{sdr} platform working in the $60$~GHz band. We tested our system on a large set of standard-compliant \ac{cir} traces matching the traffic patterns of real WiFi access points, performing a typical downstream application such as \ac{har}. Our results show that SPARCS entails over $7$ times lower overhead compared to prior methods, while achieving better performance.

Future research directions include refining SPARCS by leveraging an exchange of information between the sensing and communication processes, e.g., utilizing sensing to optimize the beam selection, and exploiting communications for collaborative sensing.  

\section*{Acknowledgments}

This research work was sponsored in part by the European Union’s Horizon 2020 research and innovation programme under grants No. 871249, on ``LOCalization and analytics on-demand embedded in the 5G ecosystem for Ubiquitous vertical applicationS'' (LOCUS) and No. 861222, on ``MIllimeter-wave NeTworking and Sensing for beyond 5G'' (MINTS), by the Spanish Ministry of Science and Innovation (MICIU) grant RTI2018-094313-B-I00 (PinPoint5G+), by the Region of Madrid through TAPIR-CM (S2018/TCS-4496) and by the Italian Ministry of Education, University and Research (MIUR) through the initiative “Departments of Excellence” (Law 232/2016). 

\bibliographystyle{ACM-Reference-Format}
\bibliography{references.bib}

\end{document}